\documentclass[prb,aps,superscriptaddress,reprint]{revtex4-2}

\usepackage{graphicx}
\usepackage{amsmath} 
\usepackage[english]{babel} 
\usepackage{units}
\usepackage[colorlinks=true,urlcolor=blue,linkcolor=blue,citecolor=blue]{hyperref}
\usepackage{mathptmx} 
\usepackage{xcolor}

\begin{document}
\title{Charge-induced artifacts in non-local spin transport measurements:\\How to prevent spurious voltage signals}

\author{Frank Volmer}
\affiliation{2nd Institute of Physics and JARA-FIT, RWTH Aachen University, 52074 Aachen, Germany}
\affiliation{AMO GmbH, Advanced Microelectronic Center Aachen (AMICA), 52074 Aachen, Germany}

\author{Timo Bisswanger}
\affiliation{2nd Institute of Physics and JARA-FIT, RWTH Aachen University, 52074 Aachen, Germany}

\author{Anne Schmidt}
\affiliation{2nd Institute of Physics and JARA-FIT, RWTH Aachen University, 52074 Aachen, Germany}

\author{Christoph Stampfer}
\affiliation{2nd Institute of Physics and JARA-FIT, RWTH Aachen University, 52074 Aachen, Germany}
\affiliation{Peter Gr\"unberg Institute (PGI-9), Forschungszentrum J\"ulich, 52425 J\"ulich, Germany}

\author{Bernd Beschoten}
\affiliation{2nd Institute of Physics and JARA-FIT, RWTH Aachen University, 52074 Aachen, Germany}

\begin{abstract}
To conduct spin-sensitive transport measurements, a non-local device geometry is often used to avoid spurious voltages that are caused by the flow of charges. However, in the vast majority of reported non-local spin valve, Hanle spin precession, or spin Hall measurements background signals have been observed that are not related to spins. We discuss seven different types of these charge-induced signals and explain how these artifacts can result in erroneous or misleading conclusions when falsely attributed to spin transport. The charge-driven signals can be divided into two groups: Signals that are inherent to the device structure and/or the measurement setup and signals that depend on a common-mode voltage. We designed and built a voltage-controlled current source that significantly diminishes all spurious voltage signals of the latter group in both DC and AC measurements by creating a virtual ground within the non-local detection circuit. This is especially important for lock-in-based measurement techniques, where a common-mode voltage can create a phase-shifted, frequency-dependent signal with an amplitude several orders of magnitude larger than the actual spin signal. Measurements performed on graphene-based non-local spin valve devices demonstrate how all spurious voltage signals that are caused by a common-mode voltage can be completely suppressed by such a current source.
\end{abstract}

\maketitle
\begin{figure*}[tb]
	\includegraphics[width=\linewidth]{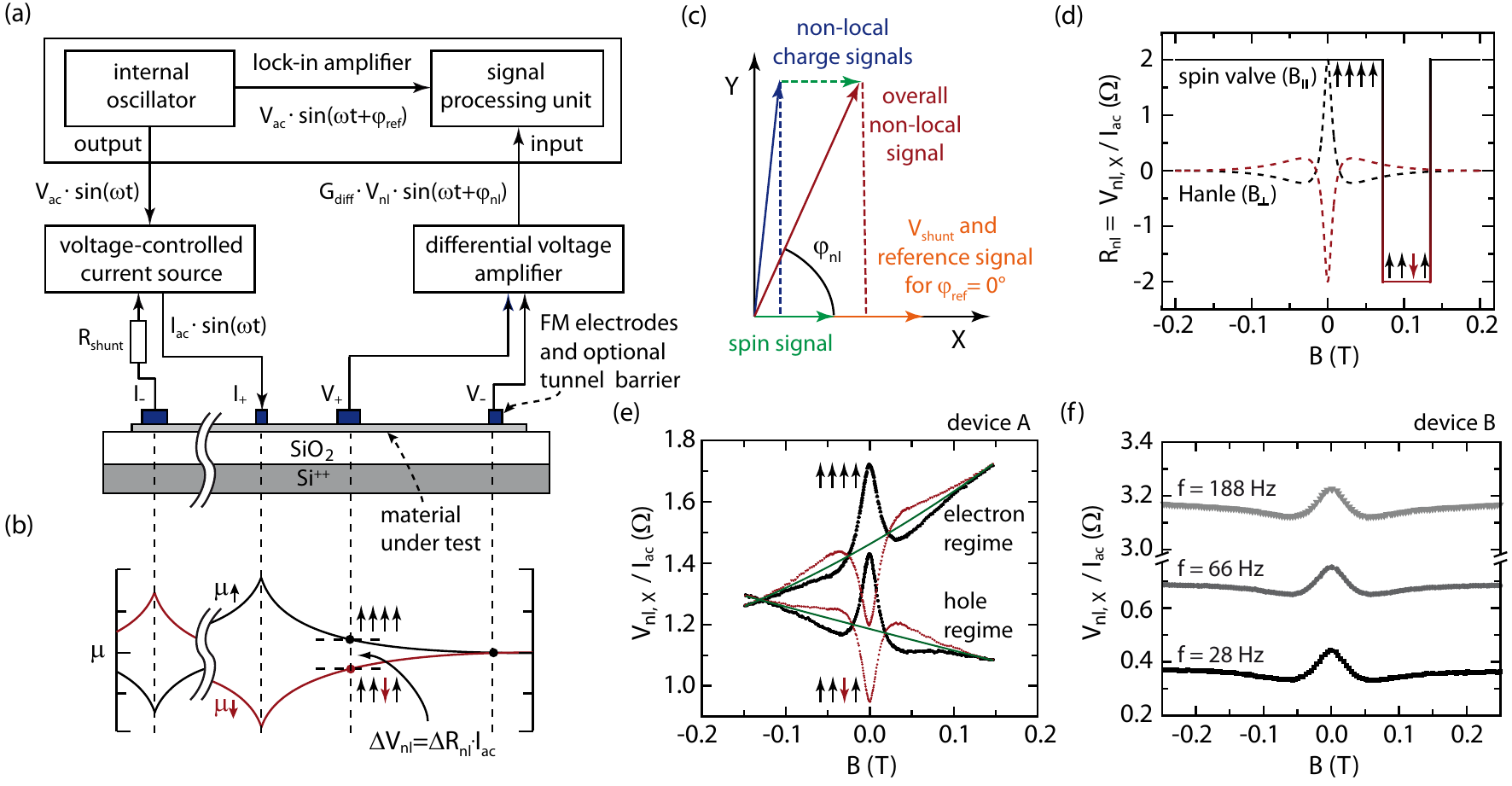}
	\caption{(a) Lock-in-based, non-local measurement scheme of a lateral spin valve device. A charge current gets spin-polarized by flowing through the ferromagnetic electrodes I$_+$ and I$_-$ and, therefore, creates the quasi-chemical potentials depicted in panel (b) for majority and minority spins. (c) As the spin signal is directly linked to the current in the injection circuit, both signals should be in phase with each other (the latter can be measured for example over a shunt resistor). As long as the voltage-to-current converter is driving the current in phase to the applied reference voltage from the lock-in, the spin signal should appear only in the X-channel of the lock-in. In real measurements also charge-induced signals appear that can be either in-phase or out-of-phase with the applied current. (d) Without these spurious signals, both spin valve (solid line) and Hanle spin precession (dashed lines) measurement should be symmetric around $R_\text{nl}=0$. Instead, in experiments a background signal is almost always measured that changes e.g.\,by going from the electron into the hole regime (e) or by merely increasing the measurement frequency (f).}
	\label{fig1}
\end{figure*}

Two of the most commonly used device geometries for spin transport measurements are non-local spin valve and non-local Hall bar geometries \cite{RevModPhys.76.323,ActaPhysSlovaca.57.565,PhysRevB.37.5312,RevModPhys.87.1213,JMagnetismMagneticMaterials.509.166711,2DMaterials.2.030202,RevModPhys.92.021003,NatureNanotechnology.16.856}. In such devices there is no spatial overlap between the injection circuit over which a charge current is driven (circuit with electrodes I$_+$ and I$_-$ in Fig.\,\ref{fig1}a) and the detection circuit in which the spin signal is non-locally probed (circuit with electrodes V$_+$ and V$_-$ in Fig.\,\ref{fig1}a). This device scheme is argued to prevent spurious voltage signals because of a putative avoidance of any charge flow in the non-local detection circuit \cite{RevModPhys.76.323,ActaPhysSlovaca.57.565,PhysRevB.37.5312,RevModPhys.87.1213}. In real devices, however, there are several mechanisms that can lead to a flow of charges in the non-local part of the device. In this article we discuss seven different mechanisms that can result in charge-induced non-local voltages. All of the discussed mechanisms are of such a fundamental nature that they can occur in any material system.

In this context, it is important to emphasize that our discussion is not exhaustive as we explicitly do not include material-specific phenomena that can lead to non-local voltage signals. One prominent example is the occurrence of large non-local signals either at the charge neutrality point in graphene or by opening a band gap in bilayer graphene. Although these non-local signals have been attributed to topological currents or the valley Hall effect \cite{Science.346.448, NaturePhysics.11.1027, NaturePhysics.11.1032, APL.114.243105}, more recent studies rather explain these measurement by generic, non-topological edge currents that lead to a flow of charges in the non-local part of the device \cite{NatureCommunications.8.2198,JournalofPhysicsMaterials.1.015006,Nature.593.528}. The ensuing debate about the actual underlying physics \cite{Nature.593.528,PhysicsWorld.November.43,JournalofPhysicsMaterials.5.021001} follows other controversial discussions such as those about the correct interpretation of non-local spin Hall effect measurements or spin transport measurements in topological insulators. For the latter two cases several studies already demonstrated the occurrence of charge-induced non-local voltages that might have been falsely attributed to spin-related effects in preceding publications \cite{PRL.103.166601,PhysRevB.99.085401,PRB.91.165412,PRL.117.176602,PRB.92.161411,PRB.92.201102}. This highlights the importance for a comprehensive review of the mechanisms that can create spurious voltage signals, which is one essential part of this article. Besides already known sources of charge-induced non-local voltages, we also discuss a spurious voltage signal that depends on the common-mode voltage in the non-local part of the device and that has not been considered so far. The amplitude of this signal scales with the applied measurement frequency. At higher frequencies this spurious voltage signal will mask any spin signal, which is the likely reason why the vast majority of reported non-local spin valve, Hanle spin precession, and spin Hall measurements were conducted with either DC currents or AC currents at very low frequencies (\unit[$\leq$30]{Hz}) \cite{NatureCommunications.6.6766,NatureCommunications.7.13372,PRL.121.127703,JApplPhysics.117.083909,NanoLett.19.59595966,Phys.Rev.Applied.10.044050,AppliedPhysicsLetters.113.132403,Phys.Rev.B.98.054412,2DMaterials.6.034003,NanoLett.19.1074,2DMaterials.6.035042,AIPAdvances.9.115005,NanoLett.16.3533,PhysRevB.90.165403,NatureMaterials.19.170175,ACSNano.14.1277112780,PRB.97.205439}. We demonstrate that both this frequency dependent spurious voltage signal and two others, which are also caused by the common-mode voltage in the detection circuit, can be completely removed by creating a virtual ground within the non-local part of the device.

In this article, we first discuss in section~\ref{Principal-of-a-non-local-spin-measurement} the fundamental principles behind non-local spin measurements to lay the basis for the discussion of the charge-induced signals. In section~\ref{Overview-of-non-local-charge-signals} we explain the mechanisms responsible for the charge-induced signals and (if available) discuss ways how their contribution to a non-local measurement can be minimized. In section~\ref{Phase-shifted-charge-signal} we discuss in detail the one charge signal whose amplitude scales with the measurement frequency and demonstrate how this signal can be removed from the measurement by using a current source that we designed and built for this purpose.

\section{Principles of a non-local spin transport measurement}
\label{Principal-of-a-non-local-spin-measurement}
We discuss the mechanisms behind the occurrence of charge-induced, non-local voltages using the example of a lateral spin-valve device whose non-local signal is measured with a lock-in technique (Fig.\,\ref{fig1}a). But we note that many of the discussed charge-induced signals also occur in other measurement techniques (including DC measurements) and other device geometries (e.g.\,Hall bar structures for the non-local detection of the spin Hall effect). In our example the sinusoidal voltage signal $V_{\text{ac}} \cdot \textrm{sin}(\omega t)$ from the internal oscillator of the lock-in amplifier is used as a reference signal for a voltage-to-current-converter. This converter is driving a current $I_{\text{ac}}\cdot \textrm{sin}(\omega t)$ through the device under test (DUT) between the two ferromagnetic electrodes in the injection circuit denoted I$_+$ and I$_-$. It is assumed that the voltage-to-current-converter only creates a negligible phase shift between the reference voltage and the current through the device (probed e.g.\,by the voltage drop over a shunt resistor $R_{\text{shunt}}$).

The applied charge current gets spin-polarized by the ferromagnetic electrodes and therefore creates a spin accumulation in the material right below the electrodes \cite{RevModPhys.76.323,ActaPhysSlovaca.57.565,PhysRevB.37.5312}. This is depicted as a quasi-chemical potential for each spin orientation in Fig.\,\ref{fig1}b. Because of the gradient in the spin accumulation, spins also diffuse from the I$_+$ electrode towards the non-local part of the device. The corresponding spin accumulation in this part decreases exponentially because of relaxation processes and can be probed between the spin-sensitive, ferromagnetic electrodes (V$_+$ and V$_-$) with a differential voltage amplifier \cite{RevModPhys.76.323,ActaPhysSlovaca.57.565,PhysRevB.37.5312}.

If the DUT has a spin diffusion coefficient $D_\text{s}$, the time that is needed for the spins to diffuse a distance $L$ into the non-local region is on the order of $L^2/D_\text{s}$ (e.g.\,for the majority of reported spin transport experiments in graphene this time is between \unit[500]{ps} and \unit[10]{ns}). In typical measurements this time is several orders of magnitude shorter than the period of the applied AC signal. Therefore, there cannot be any measurable phase that is caused by spin diffusion between the applied AC current, which creates the spin accumulation in the first place, and the measured non-local spin signal. Accordingly, any non-local signal that has a phase is either due to a measurement artifact or is linked to a physical phenomenon that is not connected to spins. In this respect a lock-in amplifier now becomes important, as it can decompose the overall non-local signal in contributions that are in-phase and out-of-phase to its internal oscillator (Fig.\,\ref{fig1}c). The X-channel (sometimes also called Re- or Real-channel) of the lock-in measures the amplitude of the non-local voltage that is in-phase to the internal oscillator, whereas the Y-channel (Im- or Imag-channel) measures the signal that is shifted by 90$^\circ$. According to the above explanation the spin signal should be entirely located within the X-channel of the lock-in (green arrow in Fig.\,\ref{fig1}c).

In Fig.\,\ref{fig1}d the theoretically expected curves for a spin valve and a Hanle spin precession measurement are depicted for the case that only the bare spin signal is present in a non-local measurement. For the spin valve measurement the magnetic field $B$ is applied in plane to the material and anti-parallel to the magnetization of the ferromagnetic electrodes. Different widths of the electrodes lead to different coercive fields because of shape anisotropy \cite{J.Appl.Phys.98.014309,PhysRevB.71.064411}. In Figs.~\ref{fig1}b and \ref{fig1}d it is assumed that electrode V$_+$ switches first (the four arrows represent the magnetisation direction going from electrode I$_-$ to V$_-$). Therefore, V$_+$ will no longer probe the spin up but rather the spin down potential, which should only lead to a sign reversal of the measured voltage (see detailed explanation with theoretical derivation in Refs.~\cite{RevModPhys.76.323,ActaPhysSlovaca.57.565,PhysRevB.37.5312}). If the non-local resistance $R_\text{nl}$ is calculated by normalizing the non-local voltage by the applied current, the corresponding spin valve curve should be perfectly centered around $R_\text{nl}=0$ (Fig.\,\ref{fig1}d). If the magnetic field $B$ is applied in the out-of-plane direction, Hanle spin precession curves can be measured both in case of a parallel orientation of the respective magnetization directions of the I$_+$ and V$_+$ electrodes (black dashed curve in Fig.\,\ref{fig1}d) and the respective anti-parallel orientation (red dashed curve). These curves are expected to be mirror images of each other and they should converge to zero at high magnetic fields because of a complete dephasing of the spins \cite{RevModPhys.76.323,ActaPhysSlovaca.57.565,PhysRevB.37.5312}.

Instead, Fig.\,\ref{fig1}e depicts typical experimental Hanle curves with obvious charge-related background signals that are measured in both the hole and the electron regime of a graphene-based spin valve device (see discussion in section~\ref{Conclusion} how the occurrence of these spurious signals may change for other materials and device structures, e.g.\,metallic spin valve devices). The background signal can be easily calculated by the arithmetic mean of the two curves for parallel and anti-parallel orientation (green curves in Fig.\,\ref{fig1}e) and can be fitted by a second order polynomial function. Subtracting this background from the measured data results in Hanle curves as depicted in Fig.\,\ref{fig1}d, i.e.\,curves that are perfectly centered around $R_\text{nl}=0$ and can be well-fitted with models that only consider a spin signal \cite{RevModPhys.76.323,ActaPhysSlovaca.57.565,PhysRevB.37.5312}. It was shown that this kind of magnetic-field dependent background signal can be explained by a combination of a current-spreading effect and the Hall effect \cite{2DMaterials.2.024001,PhysRevB.99.085401}, the latter causing the slope to change when tuning the device from the hole into the electron regime.

To highlight the previously described background signal whose amplitude increases towards higher frequencies, we use another graphene-based spin valve device for the sake of simplicity. For this device the current spreading effect was minimized by improving the device fabrication process (see explanation in section \ref{CurrentSpreading}). As a result the corresponding Hanle curves in Fig.\,\ref{fig1}f do not have a magnetic-field dependent background signal. Instead, it is seen how a constant offset voltage increases for increasing frequencies $f=\omega/(2\pi)$ of the applied current. It should be noted that in Fig.\,\ref{fig1}f only the projection of the underlying charge-induced, non-local voltage in the X-channel is depicted (compare to blue arrow in Fig.\,\ref{fig1}c). As explained in the following sections, the underlying charge-induced signal exhibits a phase close to 90$^\circ$. Accordingly, the signal in the Y-channel increases significantly more, even pushing the channel into an overload condition for frequencies higher than \unit[188]{Hz} in case of this device, which eventually leads to measurement artefacts of the spin signal in the X-channel.

There are a variety of reasons for trying to prevent such charge-induced signals from appearing in non-local measurements. The most obvious ones are that such signals can no longer be falsely attributed to spin-related effects and that the analysis of the spin data gets much simpler. Additionally, there is also the benefit to improve the signal-to-noise ratio of the spin measurement via three effects: First, the gain of the differential voltage amplifier and the sensitivity of the lock-in amplifier can be optimized for the amplitude of the spin signal. Second, preventing a flow of charges in the non-local detection circuit will also prevent that such currents can contribute noise to the measurement. Third, to be able to increase the measurement frequency opens the possibility to probe the spin signal in low-noise frequency bands far away from e.g.\,the 1/f-noise at low frequencies or typical interference frequencies of \unit[50]{Hz} or \unit[60]{Hz}.

\section{Overview of non-local charge signals}
\label{Overview-of-non-local-charge-signals}
Besides the actual spin signal $V_\text{S}$, there are at least seven other contributions to the overall non-local signal $V_\text{nl}$ that will be discussed in detail in the following sections:
\begin{equation}
V_\text{nl} = V_\text{S} + (V_\text{CS} + V_\text{T} + V_\text{IBC} + V_\text{CI}) + (V_\text{L} + V_\text{CMRR} + V_\text{CC}).
\label{AllContributionsToVnl}
\end{equation}
These contributions can be divided into two groups: The first group consists of signals that are inherent to the device structure and/or the measurement setup and are caused by current spreading ($V_\text{CS}$), thermo-electric voltages ($V_\text{T}$), input bias currents ($V_\text{IBC}$), or crosstalk and interference signals ($V_\text{CI}$). These four signals are normally quite insensitive to changes of the common-mode voltage in the transport channel, as explained in more detail in the Supporting Information \cite{Supplement}. Therefore, the current source that is presented in this study will only have minor influence on these signals. Instead, the second group consists of charge-induced signals that are caused by the common-mode voltage within the non-local detection circuit. Therefore, our current source can be used to minimize their impact on non-local spin transport measurements. This group includes signals that are caused by leakage currents ($V_\text{L}$), a finite common-mode rejection ratio ($V_\text{CMRR}$), and the signal that scales with the measurement frequency and that can be explained by charging currents of capacitances in the detection circuit ($V_\text{CC}$).

We note that the respective contribution of each charge-related signal can vary quite significantly depending on details like device fabrication, properties of the investigated material, device geometry, measurement setup, and measurement technique. Usually, there are one or two dominating charge-related signals, whereas others might only get relevant for the most precise measurements, for which voltage signals slightly above the thermal noise floor are to be measured. In section~\ref{Magnetic-field-dependence} we discuss how a magnetic field dependent measurement can be used to determine the charge-induced contribution to the total non-local signal and how this contribution can then be subtracted from the measurement to obtain the bare spin signal. Additionally, we explain possible pitfalls that may occur in such a procedure.

\begin{figure*}[tb]
	\includegraphics[width=\linewidth]{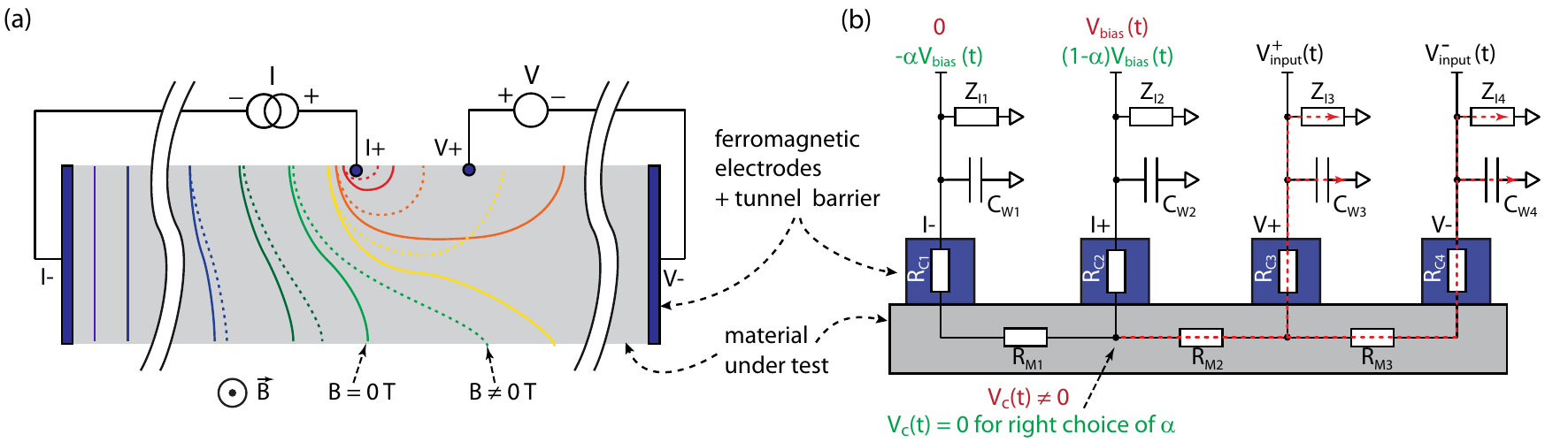}
	\caption{(a) Schematic top view of the transport channel illustrating the effect of current spreading. Because of a spatially asymmetric injection of charge carriers, a small part of the overall current flows on a curved path from I$_+$ into the non-local part of the device before it drains into I$_-$. This leads to a measurable potential difference between the two voltage probes V$_+$ and V$_-$ (colored lines illustrate equipotential lines), which can be changed by an out-of-plane magnetic field (dashed lines). (b) Simplified equivalent circuit of both the device under test and the measurement setup. Considered are the contact resistances $R_\text{Ci}$, the resistances $R_\text{Mi}$ of the material, the capacitances $C_\text{Wi}$ of the wiring, and the input impedances $Z_\text{Ii}$ of the measurement equipment. The bias voltage $V_\text{bias}(t)$ that is necessary to drive a current through the injection circuit can, in principle, be freely distributed ($0 \leq \alpha \leq 1$) between electrodes I$_+$ and I$_-$. However, there is only one value of $\alpha$, which depends on the voltage divider consisting of $R_\text{C1}$, $R_\text{M1}$ and $R_\text{C2}$, that shifts the whole non-local part of the device into a virtual ground (green letters). For every other setting of $\alpha$, a non-zero common-mode voltage $V_\text{cm}(t)\neq 0$ will drive currents over every component that is referenced to ground in the non-local part of the device (red arrows).}
	\label{fig2}
\end{figure*}

\subsection{Current Spreading ($V_\text{CS}$)}
\label{CurrentSpreading}
One of the most documented charge signals unfortunately does not have a consistent name convention, but instead can be found by different names in literature, such as "current spreading", "baseline resistance", or simply "ohmic contribution" \cite{PhysRevB.76.153107,2DMaterials.2.024001,PRL.103.166601,PhysRevB.99.085401,IEEETransactionsonElectronDevices.66.50035010,APL.99.142112,AIPAdvances.9.115005}. Common to all publications about this charge signal is a spatially inhomogeneous, non-uniform injection of the charge current into the DUT, either via a side-arm of a Hall bar structure or a pinhole in the tunnel barrier of the injection electrode. This is illustrated in Fig.\,\ref{fig2}a in which I$_-$ drains the current and V$_-$ detects any voltage uniformly over the whole width of the DUT. Electrode I$_+$ instead injects the current in a point-like manner, breaking the spatial symmetry. A small part of the sourced current will flow on a curved path from I$_+$ into the non-local part of the device before it eventually flows into I$_-$. This leads to a non-uniform potential landscape in the non-local part of the device (colored lines illustrate equipotential lines; actual simulations of this effect can be found in Refs.~\cite{2DMaterials.2.024001,PhysRevB.99.085401,AIPAdvances.9.115005,2DMaterials.9.015024}). This non-uniform potential landscape results in a potential difference $V_\text{CS}$ between the two voltage probes V$_+$ and V$_-$ (especially if V$_+$ also probes in a point-like manner).

An out-of-plane magnetic field (as applied in a Hanle spin precession measurement) acts on the charge currents via a Lorentz force that modifies the potential landscape and therefore the measured non-local voltage (an illustration of the resulting change in the potential landscape is shown by dashed lines in Fig.\,\ref{fig2}a). Numerical simulations have shown that this Hall-like effect can explain the background signal in Fig.\,\ref{fig1}e that depends both on the magnetic field and the charge carrier density \cite{2DMaterials.2.024001,PhysRevB.99.085401}. In case of a Hall bar geometry, the magnitude of the non-local charge signal at $B=\unit[0]{T}$ can be estimated by the van der Pauw theorem to \cite{PRL.103.166601}:
\begin{equation}
    V_\text{CS} = I \cdot\rho\cdot \exp\left(-\pi \frac{L}{w}\right),
    \label{Eq:CS}
\end{equation}
with the applied current $I$, the two-dimensional sheet resistance $\rho$, the distance $L$ between the sidearm at which the current is injected and the sidearm at which the non-local voltage is probed, and the width $w$ of the transport channel.

We are not aware of a measurement technique that can minimize the contribution of $V_\text{CS}$ to the overall non-local signal. Nevertheless, certain changes in the fabrication process can minimize this charge signal. Foremost, according to equation~\ref{Eq:CS} an increase in the aspect ratio $L/w$ can significantly reduce the effect of current spreading both in case of Hall bar and spin valve geometries. As both the signal due to current spreading and the spin signal decay exponentially with the distance $L$ between the I$_+$ and V$_+$ electrodes \cite{RevModPhys.76.323,ActaPhysSlovaca.57.565,PhysRevB.37.5312}, it is especially the width $w$ of the transport channel that should be minimized. Only applicable to the spin valve geometry is the improvement of the interface between the ferromagnetic electrode and the material under test (e.g.\,by minimizing pinholes in a tunnel barrier), to increase the uniformity of the charge carrier injection over the whole width of the transport channel.

\subsection{Thermoelectric voltages ($V_\text{T}$)}
It is well-documented that thermo-electric and even magneto-thermo-electric effects can cause non-local charge signals via a combination of Joule heating, the Peltier effect, the Seebeck effect, the Ettingshausen effect, or the Nernst effect \cite{PRL.112.116601,PRL.105.136601,PRB.94.180403,NatureComm.6.8337,PRB.87.014424,PRB.100.104404,IEEETransactionsonElectronDevices.66.50035010,AppliedPhysicsExpress.10.063004}.
The heat that is generated by the flow of charge carriers between the injection electrodes I$_+$ and I$_-$ (Fig.\,\ref{fig1}a) creates a temperature gradient along the non-local part of the device, especially as electrodes V$_+$ and V$_-$ can act as heat sinks. This temperature gradient then results in a differential voltage $V_\text{T}$ via e.g.\,the Seebeck or Nernst effect \cite{PRL.112.116601,PRL.105.136601,PRB.94.180403,NatureComm.6.8337,PRB.87.014424,PRB.100.104404,AppliedPhysicsExpress.10.063004}.

The heat that is created by the flow of charge carriers in the injection circuit can depend both linearly (e.g.\,in case of the Peltier effect) or quadratically (e.g.\,in case of Joule heating) on the applied current. Hence, in a lock-in measurement the resulting thermoelectric signal can appear both at the first and the second harmonic of the excitation frequency of the current \cite{PRL.112.116601,PRL.105.136601}. To our knowledge, no phase shift is observed between the applied current and the thermally induced voltage. This is consistent to the thermal diffusivity of typical materials that are used for spin transport devices \cite{AIPAdvances.2.041410,NatureMaterials.10.569}. These thermal diffusivities yield heat transfer times over the whole device that are several orders of magnitude shorter than the period of typically applied AC signals (see similar discussion about the spin diffusion coefficient in section~\ref{Principal-of-a-non-local-spin-measurement}).

Because of the inherent nature of thermoelectric voltages, we are not aware of measurement methods that can reduce their contribution after device fabrication. A suppression of these spurious signals can only occur by design choices during the fabrication process, like the design of electrodes that may function as heat sinks, the heat conductivity of the underlying substrate, or the design of the I$_+$ and I$_-$ electrodes in such a way that their respective Peltier effects cancel each other to some extent \cite{PRB.100.104404,IEEETransactionsonElectronDevices.66.50035010,AppliedPhysicsExpress.10.063004}.

Finally, it is important to note that thermoelectric signals not only depend on gate-induced changes in the transport properties of the DUT \cite{PRL.102.096807,2DMaterials.7.041004}, but that there are even magnetic field dependent analogs to the Peltier and Seebeck effects, whiche are the Ettingshausen and Nernst effects, respectively. These two effects can result in a $B$-field dependence of the corresponding non-local, thermoelectric signal \cite{PRL.112.116601,NatureComm.6.8337,PRB.94.180403}.

\subsection{Input bias currents ($V_\text{IBC}$)}
\label{InputBiasCurrents}
We now assume that no current source is connected to the device and that the electrodes I$_+$ and I$_-$ are shorted to GND. The only equipment that is connected to the DUT via electrodes V$_+$ and V$_-$ is either a differential voltage amplifier or a lock-in amplifier. But even in this situation a small current will flow through the device. This is due to currents that are flowing either out of or into the inputs of a voltage or a lock-in amplifier. Especially in case of operational amplifiers or instrumentation amplifiers this current is called input bias current \cite{AnalysisAndDesignOfAnalogIntegratedCircuits,IEEEJournalofSolid-StateCircuits.17.969,IEEEJournalofSolidStateCircuits.40.1212}. In a transport measurement these input bias currents result in non-local charge currents and, therefore, generate voltage drops $V_\text{IBC}$ within the device.

The input bias currents differ by several orders of magnitude between different measurement equipments. Because of high input bias currents up to the nA or even $\mu$A range, standard amplifiers that use bipolar junction transistors (BJT) in their input stages are unsuitable for non-local spin measurements. However, precision amplifiers may include internal circuitry that diminishes the input bias currents of BJT inputs down to the higher pA range. Even lower input bias currents in the lower pA or even fA range \cite{RevSciInst.88.085106} can be found in amplifiers that use field-effect transistors (FET). However, even this type of amplifier might cause problems, as a damage to the input stages may increase these currents significantly.

It is a-priori not clear if the charge signal that is caused by input bias currents is relevant in a non-local spin measurement. We recommend to measure both the input bias currents and the the common-mode rejection ratio $CMRR$ (see section~\ref{CommonModeRejectionRatio}) of the experimental setup to clarify this point. In the Supplemental Material \cite{Supplement} we give more information about such measurement procedures and discuss why input bias currents are normally only relevant for DC measurements.

\subsection{Crosstalk and interference signals ($V_\text{CI}$)}
\label{CrosstalkAndInterference}
There is always a certain coupling between different signal lines within a measurement setup. Especially cryogenic measurement systems, for which unshielded cables are often placed close together over longer distances \cite{CryostatDesign}, can exhibit non-negligible capacitive coupling between individual lines. An AC signal, like the bias voltage that drives the current through the injection circuit, can therefore couple into the wires that are connecting the V$_+$ and V$_-$ electrodes to the measurement equipment outside the cryostat. Everything that will change the amplitude of the bias voltage, like a gate-induced change in the conductance of the DUT or a magnetoresistance effect, can change the amplitude of the signal $V_\text{CI}$ that is coupled into the non-local detection circuit.

Wires can also be prone to damages to their insulation, which can yield reduced insulation resistances between different lines. Leakage currents directly between wires of the injection and detection circuit can therefore result in spurious signals in case of both AC and DC measurements. Fortunately, such setup-related signals can be easily identified by measuring the impedances between each line without any installed DUT.

Interference signals that couple from external sources into the measurement setup can also result in spurious, non-local signals. A common but erroneous assumption is that radio frequency (RF) signals do not impact DC or lock-in measurements, as in the first case the signal is averaged over time-scales several orders of magnitude longer than the period of the RF signal and in the second case the lock-in only measures a signal within a very narrow band around the excitation frequency. But as long as there are non-linear components within the measurement circuit, a down-mixing of the RF interference signal occurs. As a result, RF signals are known to e.g.\,create DC offset voltages in amplifier circuits \cite{IEEETransactionsonCircuitsandSystems.49.367,ElectronicsLetters.30.282,ElectronicsLetters.43.1088}.

\subsection{Common-mode voltage and virtual ground}
\label{CommonModeVoltageAndVirtualGround}
All of the non-local charge signals, which are discussed in the following sections, significantly depend on the so-called common-mode voltage $V_\text{cm}(t)$ in the non-local part of the device. This is the voltage that is common to both inputs (non-inverting (+) and inverting (-) input) of the differential amplifier that is connected to electrodes V$_+$ and V$_-$. If we denote these input voltages as $V_\text{input}^{+/-}$, then the common-mode voltage is defined as:
\begin{equation}
    V_\text{cm}(t) = \frac{V_\text{input}^{+}(t)+V_\text{input}^{-}(t)}{2}.
\end{equation}
To simplify the discussion on this matter, we make several assumptions that should apply well to a majority of devices. However, these assumptions should be carefully examined considering own device schemes and measurement techniques.

We assume that the frequency of the applied AC current is low enough that any inductance or capacitance of the DUT can be neglected, while the measurement setup can still exhibit inductances or capacitances. The analog assumption in case of a DC measurement is that the time step between two measurement points is much longer than the settling time that is caused by the inductance or capacitance of the DUT. Accordingly, we model the equivalent circuit of the device by a series of ohmic resistors (Fig.\,\ref{fig2}b) where the contact resistances $R_\text{Ci}$ exhibit ohmic characteristics. Finally, we assume that the measurement system exhibits a well-defined ground potential (GND), that every voltage is referenced to this GND and that the shielding of any coaxial cables are put to the same GND.

To drive a current $I_\text{bias}$ through the injection circuit, a current source has to apply the voltage
\begin{equation}
V_\text{bias} = I_\text{bias}\cdot\left( R_\text{C1} + R_\text{M1} + R_\text{C2} \right)
\end{equation}
between electrodes I$_+$ and I$_-$. The easiest way to accomplish this is to put one of the electrodes to GND and to apply the full bias voltage (referenced to GND) to the other electrode (red letters in Fig.\,\ref{fig2}b). Because of the voltage divider that is built from $R_\text{C1}$, $R_\text{M1}$ and $R_\text{C2}$, the material under test right below the injection electrode I$_+$ will be on a potential $V_\text{c}(t)$ that is unequal to zero, i.e.\,unequal to GND. Every measurement parameter that has an effect on one of the resistances $R_\text{C1}$, $R_\text{M1}$ and $R_\text{C2}$ will thus have an impact on $V_\text{c}(t)$. E.g.\,changing the conductivity of the material by tuning the charge carrier density via a gate voltage will change $R_\text{M1}$. Therefore, all spurious non-local signals that are caused by $V_\text{c}(t)$ will show a gate-dependent behaviour.

If we neglect any charge currents and their associated ohmic voltage drops in the non-local part of the device, the same voltage $V_\text{c}(t)$ is applied to both inputs of the differential voltage amplifier that is connected to electrodes V$_+$ and V$_-$. Under this condition the common-mode voltage $V_\text{cm}(t)$ would be equal to $V_\text{c}(t)$.

As explained in the following sections, several non-local signals are caused by this common-mode voltage. Fortunately, there is a quick way to check if $V_\text{cm}$ is playing a role in a measurement. For the wiring denoted with the red letters in Fig.\,\ref{fig2}b, the measurement only has to be repeated after putting a high-ohmic resistor between the I$_-$ electrode and GND. The current-induced voltage drop over this additional resistor will lift the potential (measured against GND) in the whole circuit, therefore increasing $V_\text{cm}$. There is only one pitfall to this method, which is a general problem for non-zero common-mode voltages in devices in which the charge carrier density is controlled via electrostatic gating: Gate voltages are normally referenced to GND, but the gate-electric field effect does not depend on the absolute value of the gate voltages but the potential differences between the gates and the DUT. Changing $V_\text{cm}$ is therefore also changing the gate-induced charge carrier densities (see Supplemental Material \cite{Supplement} for more information). If the spin-signal depends on the charge carrier density, it might be difficult to tell if a change in the non-local signal by lifting the non-local part of the DUT to another potential is related to either a change in the charge-carrier dependent spin signal or a change in the common-mode dependent spurious signals.

The current source that is presented in this work can significantly reduce the common-mode voltage by freely distributing the potential difference that is necessary to drive the current to the two injection electrodes (green letters in Fig.\,\ref{fig2}b):
\begin{equation}
 V_\text{bias}(t) = \left[ (1-\alpha)\cdot V_\text{bias}(t)\right] - \left[ -\alpha\cdot V_\text{bias}(t) \right],
\end{equation}
with $\alpha$ adjustable between $0 \leq \alpha \leq 1$ (see in-depth technical discussion of the current source in the Supplemental Material \cite{Supplement}). Depending on the values of the ohmic voltage divider consisting of $R_\text{C1}$, $R_\text{M1}$ and $R_\text{C2}$, there will be a certain value of $\alpha$ for which the common-mode voltage is zero ($V_\text{cm}(t) = 0$). Although the non-local part of the device is not connected to GND, it will nevertheless be on ground potential at all times. The right choice of $\alpha$ will therefore create a so-called virtual ground in the non-local part of the device. This not only nullifies the non-local charge signals that are caused by $V_\text{cm}$, but also results in the fact that any applied gate voltage (referenced to GND) is equal to the potential difference between gate and the non-local part of the DUT (see Supplemental Material \cite{Supplement} on how the adjustment of $\alpha$  therefore also impacts a local four-probe measurement of the gate-dependent resistivity of graphene).

\subsection{Voltages caused by leakage currents ($V_\text{L}$)}
\label{LeakageCurrents}
The most obvious non-local charge signal that is caused by a common-mode voltage $V_\text{cm}$ is due to leakage currents to GND. Especially the finite input impedances of measuring equipment ($Z_\text{Ii}$ in Fig.\,\ref{fig2}b) play a crucial role, but also more unexpected current paths to GND like damaged insulations of cables or broken capacitors. These leakage currents (red, dashed lines in Fig.\,\ref{fig2}b) create ohmic voltage drops over the resistances $R_\text{C3}$, $R_\text{M3}$ and $R_\text{C4}$, which result in a non-local voltage $V_\text{L}$ between V$_+$ and V$_-$.

Assuming purely ohmic impedances and a voltage divider model, the amplitude of this non-local voltage scales linearly with the common-mode voltage $V_\text{cm}$ and significantly depends on the relative magnitude of the leakage resistance to ground with respect to the resistances $R_\text{C3}$, $R_\text{M3}$ and $R_\text{C4}$. Therefore, it is not surprising that spurious voltage signals due to leakage currents are mainly reported in measurements of high-impedance states (see Ref.~\cite{RevSciInst.89.024705} and the supplementary information of Refs.~\cite{NaturePhysics.11.1027,NaturePhysics.11.1032}). Hence, the spurious signal $V_\text{L}$ can e.g.\,be much more pronounced in graphene-based spin valves, in which high-resistive tunnel barriers have to be incorporated to overcome the conductivity mismatch problem \cite{PhysRevB.62.R16267,PhysRevB.62.R4790,PhysRevB.64.184420} compared to metallic spin valves without tunnel barriers.

In our work on graphene-based spin valves, we use a lock-in amplifier that has an input impedance of only $\unit[10]{M\Omega}$ to GND (SR830 from Stanford Research Systems). To reduce the effect of leakage currents, we use an additional differential amplifier (SR560 from Stanford Research Systems) with a higher input impedance of $\unit[100]{M\Omega}$ to probe the differential voltage between electrodes V$_+$ and V$_-$ before the non-local signal is sent to the input of the lock-in (Fig.\,\ref{fig1}a). Some laboratory-grade multimeters and voltmeters even offer the possibility to select the input impedance to be either $\unit[10]{M\Omega}$ or $\unit[10]{G\Omega}$. Repeating a non-local measurement with different input impedances can easily answer the question if the measured non-local voltage is partially caused by leakage currents. But the best solution is to avoid leakage currents entirely by creating a virtual ground inside the non-local part of the device. Under such conditions, the absence of a potential difference between the non-local part of the device and GND completely prevents leakage current from flowing.

\subsection{Common mode rejection ratio ($V_\text{CMRR}$)}
\label{CommonModeRejectionRatio}
The output voltage $V_\text{out}^\text{ideal}$ of an ideal differential voltage amplifier, which has a differential gain of $G_\text{diff}$ and is connected to electrodes V$_+$ and V$_-$ (Fig.\,\ref{fig1}a), is given by:
\begin{equation}
V_\text{out}^\text{ideal}=G_\text{diff}\cdot\left(V_\text{input}^{+} - V_\text{input}^{-}\right).
\end{equation}
However, the output voltage $V_\text{out}^\text{real}$ of a real differential amplifier also contains a contribution that is proportional to the average voltage at electrodes V$_+$ and V$_-$ and that scales with the common-mode gain $G_\text{cm}$ \cite{AnalysisAndDesignOfAnalogIntegratedCircuits,IEEEJournalofSolid-StateCircuits.17.969,ElectronicsLetters.19.547,IEEETransactionsonInstrumentationandMeasurement.40.669,RevSciInst.89.024705,RevSciInst.88.085106}:
\begin{equation}
    V_\text{out}^\text{real} = G_\text{diff}\left( V_\text{input}^{+} - V_\text{input}^{-} \right) + G_\text{cm} \frac{\left( V_\text{input}^{+} + V_\text{input}^{-} \right)}{2}.
\end{equation}
Neglecting any current-driven voltage drops in the non-local part of the device, the average voltage of electrodes V$_+$ and V$_-$ is exactly the voltage that is denoted as the common-mode voltage $V_\text{cm}$ in section~\ref{CommonModeVoltageAndVirtualGround}. Therefore, $V_\text{cm}$ will appear in the non-local measurement, although strongly attenuated by the common-mode gain $G_\text{cm}$. Data sheets of measurement equipment often state the ratio between the differential and the common-mode gain, the so-called common mode rejection ratio ($CMRR$):
\begin{equation}
CMRR\left[\frac{\mu\text{V}}{\text{V}}\right] = \frac{G_\text{diff}}{G_\text{cm}},\;
CMRR\left[\text{dB}\right] = 20\cdot\log\left(\frac{G_\text{diff}}{G_\text{cm}}\right).
\end{equation}
In some data sheets the $CMRR$ is only given for a specific differential gain and measurement frequency. As the $CMRR$ can depend quite significantly on these two parameters, it is always a good practice to measure the $CMRR$ of the own setup at gains and frequencies used for the experiments.

\subsection{Capacitor charging ($V_\text{CC}$)}
\label{CapacitorCharging}
All previously discussed signals are in phase with the current that is flowing in the injection circuit and are thus in phase with the spin signal. In contrast, the charge signal that is discussed in this section is phase-shifted to values close but unequal to 90$^\circ$. If a sinusoidal current is driven through the injection circuit, also $V_\text{cm}(t)$ and, therefore, the whole potential of the non-local part of the device will change sinusoidally over time. This is a problem as there are capacitances to GND in the non-local detection circuit ($C_\text{W3}$ and $C_\text{W4}$ in Fig.\,\ref{fig2}b). Some of these capacitances are intentional, like the capacitors of RC low-pass filters, others are either parasitic or unavoidable, like capacitances to GND of the wiring (around \unit[100]{pF/m} in case of coaxial cables) or the input capacitance of measurement equipment. As a result, the time-varying common-mode voltage $V_\text{cm}(t)$ will drive charge and discharge currents over these capacitors (dashed red lines in Fig.\,\ref{fig2}b).

The resulting differential voltage $V_\text{CC}$ between V$_+$ and V$_-$ can be best understood, if the combinations of the device resistances ($R_\text{Ci}$ and $R_\text{Mi}$) and the setup capacitances ($C_\text{Wi}$) are seen as RC low-pass filters for the voltage $V_\text{c}(t)$ in Fig.\,\ref{fig2}b. The respective RC time constants of the low-pass filters of the voltage probes V$_+$ and V$_-$ typically differ, i.e.:
\begin{equation}
    \left(R_\text{M2}+R_\text{C3}\right)\cdot C_\text{W3}\neq \left(R_\text{M2}+R_\text{M3}+R_\text{C4}\right) \cdot C_\text{W4}.
\end{equation}
As a result, both the amplitude and the phase of the transmitted voltage signal will differ between the two inputs of the differential voltage amplifier. This leads to a partial conversion of the common-mode voltage into a differential voltage, which is a well-known phenomenon in electronics. Such a conversion is discussed in many application notes of analog-to-digital converters, where a mismatch in the capacitors of a common-mode filter can convert common-mode noise into differential noise. Another example is the conversion between common- and differential-mode voltages caused by imbalances in filters against electromagnetic interference \cite{IEEETransactionsOnElectromagneticCompatibility.52.578,ElectromagneticCompatibilityEngineering}.

Although the voltage $V_\text{CC}$ is caused by the charging currents of the capacitors $C_\text{W3}$ and $C_\text{W4}$, it is nevertheless not perfectly at a phase of 90$^\circ$ relative to $V_\text{c}(t)$ due to the ohmic resistances $R_\text{M2}$, $R_\text{M3}$, $R_\text{C3}$ and $R_\text{C4}$. Therefore, $V_\text{CC}$ also provides a contribution to the X-channel signal of the lock-in amplifier, which will be discussed in  section~\ref{Phase-shifted-charge-signal} and can also be seen in simulations conducted with LTspice (models used for the simulations are available in Ref.~\cite{Zenodo}).

The amplitude $M=\sqrt{X^2+Y^2}$ of $V_\text{CC}$ scales linearly with the measurement frequency (see section~\ref{Phase-shifted-charge-signal}), because the current $I_\text{cap}$ that is caused by a voltage $V_\text{cap}(t)=V_\text{cap}^0\cdot\sin(\omega t)$ over a capacitor $C$ is given as:
\begin{equation}
    I_\text{cap}(t) = C \frac{\text{d} V_\text{cap}(t)}{\text{d} t} = \omega C  V_\text{cap}^0\cdot\cos(\omega t).
\end{equation}
This capacitor charging current also flows through the non-local part of the device. In general, $V_\text{CC}$ is highly dependent on every change in the device's resistances and capacitances: A change in $R_\text{C1}$, $R_\text{M1}$, or $R_\text{C2}$ will change the value of $V_\text{c}(t)$ and, therefore, also the voltage that will be applied to the capacitors $C_\text{W3}$ and $C_\text{W4}$. On the other hand, every change in $R_\text{M2}$, $R_\text{M3}$, $R_\text{C3}$, $R_\text{C4}$, $C_\text{W3}$, or $C_\text{W4}$ will have an impact on the RC time constants.

To minimize $V_\text{CC}$, the capacitances $C_\text{W3}$ and $C_\text{W4}$ should be minimized as much as possible, e.g.\,by reducing the length of the wiring from the V$_+$ and V$_-$ electrodes or by avoiding any RC low-pass filters in the measurement setup. Furthermore, additional ohmic resistors might be put into series to the electrodes V$_+$ and V$_-$ to roughly match the RC time constants by compensating varying contact resistances. But the best way to suppress $V_\text{CC}$ is to set the whole non-local part of the device to a virtual ground. In such a case, there is no potential difference across the capacitors $C_\text{W3}$ and $C_\text{W4}$ and, therefore, no capacitor charging current.

\stepcounter{subsection}
\subsection{Magnetic field dependent measurements}
\label{Magnetic-field-dependence}
Magnetic field dependent transport measurements are usually performed to distinguish an actual spin signal from spurious charge signals. The success of such an approach significantly depends on two conditions:
\begin{enumerate}
    \item Does the spin signal exhibit a magnetic field dependence that significantly differs from the magnetic response of the charge-induced signals? The best case scenario here is the existence of a clear spin precession signal.
    \item Does the experiment allow to switch the orientation of the injected spins, e.g.\,by switching the magnetization of the ferromagnetic electrodes?
\end{enumerate}
If both conditions are met, the most reliable way to mathematically remove any charge-induced contribution from the measured data is to record the Hanle spin precession curves for both parallel ($\uparrow \uparrow$) and anti-parallel ($\uparrow \downarrow$) orientations of the respective magnetization of the injection and detection electrodes. In the ideal case only the bare spin signal $V_\text{S}$ will change its sign when reversing the relative magnetization directions \cite{RevModPhys.76.323,ActaPhysSlovaca.57.565}. Instead all other spin-independent, charge-induced signals in equation~\ref{AllContributionsToVnl} should not change with the magnetization of the electrodes (exceptions are discussed further below). Therefore, the arithmetic mean of both measurements should yield the charge-induced background signal:
\begin{equation}
    V_\text{background} = \frac{V_{\text{nl},\uparrow \uparrow}+V_{\text{nl},\uparrow \downarrow}}{2}.
    \label{BackgroundDetermination}
\end{equation}
This is shown as the green line in Fig.\,\ref{fig1}e. Subtracting both measurements from each other should yield the pure spin signal:
\begin{equation}
    2\cdot V_\text{S} = V_{\text{nl},\uparrow \uparrow}-V_{\text{nl},\uparrow \downarrow}.
    \label{BackgroundSubstraction}
\end{equation}
Under such conditions the charge-induced signals should not pose any risk of a misinterpretation of the spin-dependent data. Nevertheless, the noise that is associated with the charge-induced signals cannot be removed by this subtraction method and, therefore, will deteriorate the signal-to-noise ratio of the spin signal.

It is important to note that equations~\ref{BackgroundDetermination} and \ref{BackgroundSubstraction} are no longer valid if the ferromagnetic electrodes exhibit fringe or stray fields that can act on the DUT. Such fields are well known to complicate the analysis of spin measurements \cite{PRB.92.201102,2DMaterials.5.014001,PhysRevApplied.15.044010,J.Phys.D.AppliedPhysics.49.133001,PRB.84.054410,Nanotechnology.24.015703,2DMaterials.9.015024} and can appear in quite different forms:
1.) As long as the width of the electrode is small enough that a single-domain magnetization prevails \cite{PhysRevB.71.064411}, fringe fields with relevant strengths should only be located at the electrode's two faces that are normal to the electrode's easy-axis \cite{PhysRevB.71.064411,APL.107.142406}. Stray fields can impact the measurement if these two faces are placed too close to or even on top of the transport channel \cite{PRB.92.201102,2DMaterials.5.014001,PhysRevApplied.15.044010}.
2.) If the width of the electrode is too wide, the magnetization can fall into a multi-domain structure that exhibits stray fields all along its different faces \cite{PhysRevB.71.064411}.
3.) A magnetic field applied non-colinear to the electrode can rotate the magnetization from the electrode's easy axis \cite{J.Appl.Phys.98.014309}, possibly pushing fringe fields towards the transport channel. This can lead to a situation where fringe fields increasingly affect the DUT with increasing external magnetic field strength (see main text and supplementary information of Refs.~\cite{NanoLett.19.59595966,ACSNano.14.5251,2DMaterials.9.015024}).
4.) Stray fields can also occur if the surface of the ferromagnet exhibits a great roughness, which e.g.\,can be caused by pinholes in a tunnel barrier that is separating the ferromagnet from the DUT \cite{J.Phys.D.AppliedPhysics.49.133001,PRB.84.054410,Nanotechnology.24.015703}.
5.) Stray fields can also be created by domain walls that are pinned at steps in the ferromagnetic electrode. Even the small height difference between a substrate and the top of a single layer graphene spin valve was found to be large enough to be able to pin a domain wall \cite{APL.107.142406}.

\begin{figure*}[tb]
	\includegraphics[width=\linewidth]{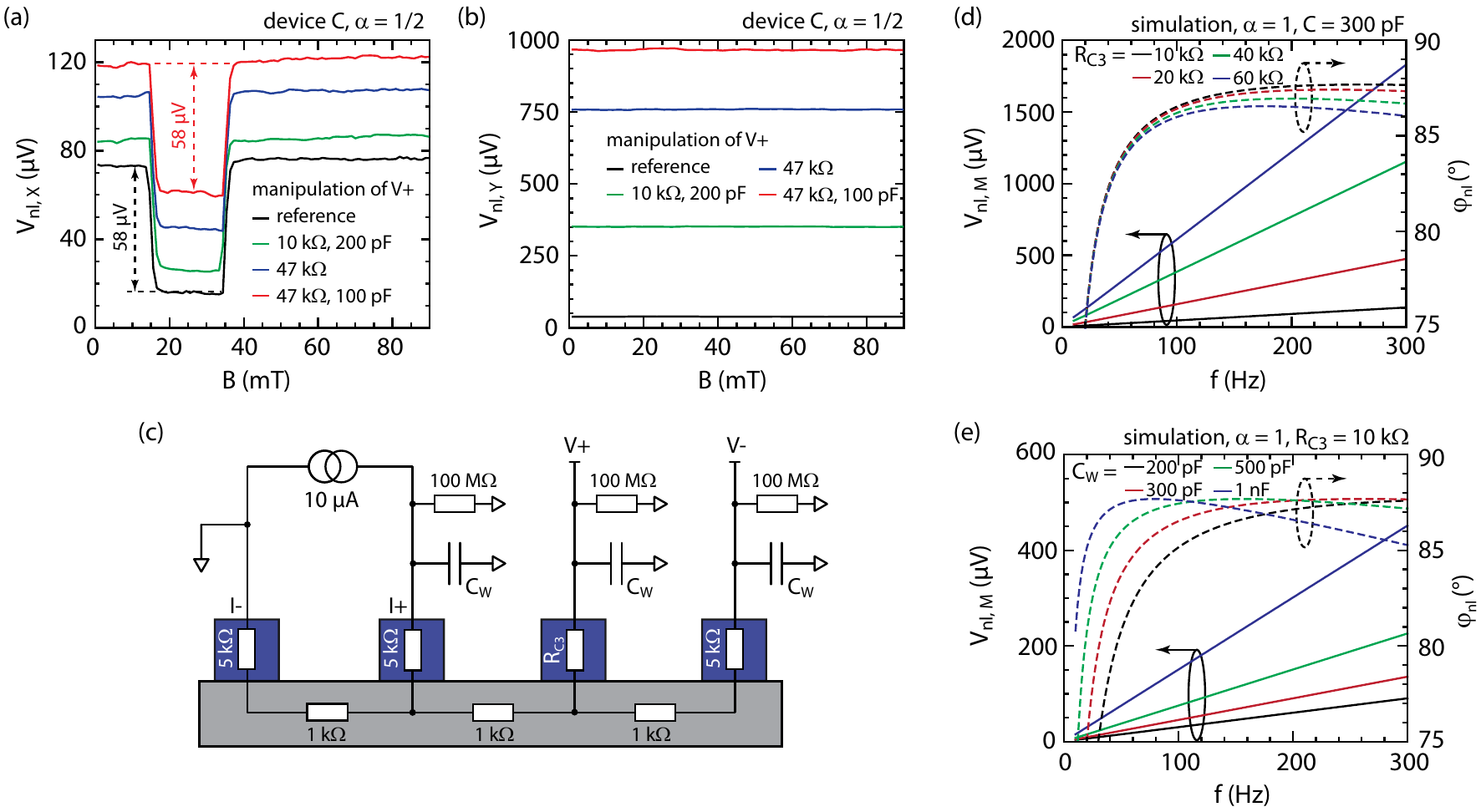}
	\caption{Demonstration on how changing the RC time constant of the V$_+$ electrode impacts the measured non-local voltage. (a), (b): Spin valve measurements are repeated with additional resistors and BNC cables (\unit[100]{pF/m}) connected to the V$_+$ electrode. The impact on the recorded non-local signal is depicted in panel (a) for the X-channel and in panel (b) for the Y-channel of the lock-in. (c) Equivalent circuit that is used to simulate the effect of varying RC time constants. (d) Corresponding amplitude $M=\sqrt{X^2+Y^2}$ and phase of the non-local voltage measured between V$_+$ and V$_-$ if the resistance $R_\text{C3}$ in (c) is varied for a fixed capacitance $C_\text{W}$ of the external wiring. (e) Same as in panel (d) but for a fixed value $R_\text{C3}$ and different values for $C_\text{W}$.}
	\label{fig3}
\end{figure*}

Fringe or stray fields can become a real issue if there is no unmistakable signature of spin precession in a magnetic field dependent measurement. For example, in 3D topological insulators (TIs) spin precession in a non-local spin transport measurement is not expected  because of the spin-momentum-locking \cite{RevModPhys.82.3045}. This is the reason why only spin valve measurements are shown in a large number of publication about spin transport experiments in TIs \cite{NatureNanotechnology.9.218,NanoLett.14.5423,NanoLett.14.6226,NanoLett.15.7976,ScientificReports.5.14293,PRB.92.155312,PRB.94.075304}. In this context, it was indeed demonstrated that fringe-field-induced Hall voltages can create artefacts that resemble the switching signals that are expected from a current-induced spin polarization in TI-based spin valves \cite{PRB.92.201102,2DMaterials.5.014001,Nanotechnology.24.015703}.

If neither of the two above conditions is fulfilled, it is even more important to carefully analyze the non-local spin measurements to avoid a misinterpretation of the results. This is especially the case for non-local spin Hall experiments, where quite controversial results were published over the years \cite{RevModPhys.87.1213}. In particular, reports about unexpectedly large spin Hall effects \cite{NatureMaterials.7.125,NaturePhysics.9.284,NatureComm.5.4748} often could not be reproduced in later studies that instead identified the role of charge-induced signals in the non-local measurement \cite{PRL.103.166601,PhysRevB.99.085401,PRB.91.165412,PRL.117.176602,PRB.92.161411}.

Unfortunately, many studies only consider the one charge-induced signal that contributes the most to the overall non-local signal. After subtracting this spurious signal from the measured data, it is then argued that the remaining signal has to be due to spins, especially when the signal shows a magnetic field dependence. However, quite a few charge-induced signals can also exhibit a magnetic field dependence: All charge currents that are flowing in the non-local part of the device, either caused by current spreading or leakage currents, experience the Hall effect. Furthermore, thermal voltages due to the Ettingshausen and Nerst effects show a magnetic field dependence. And finally, any magnetoresistive effect that changes the resistance of the DUT in the injection circuit ($R_\text{M1}$ in Fig.\,\ref{fig2}b) will have an impact on the common-mode voltage $V_\text{cm}$ and, therefore, will change all charge-induced signals that depend on $V_\text{cm}$, i.e. $V_\text{L}$ (section~\ref{LeakageCurrents}), $V_\text{CMRR}$ (section~\ref{CommonModeRejectionRatio}), $V_\text{IBC}$
(section~\ref{InputBiasCurrents}), and $V_\text{CC}$ (section~\ref{CapacitorCharging}). The magnetic-field-induced changes might be small, but so are the majority of measured non-local spin signals.

\section{Phase-shifted charge signal in lock-in measurements}
\label{Phase-shifted-charge-signal}
In this section, we discuss in detail the charge-induced non-local voltage signal that is caused by the charging and discharging currents of the capacitances in the non-local part of the setup (see section~\ref{CapacitorCharging}). For this purpose, we measure graphene spin-valve devices with cobalt electrodes and either MgO or Al$_2$O$_3$ tunnel barriers (details on the exact device geometry and the fabrication process including flake transfer techniques, etching recipes in case of structured CVD-graphene devices, and metalization schemes can be found in Refs.~\cite{NanoLett.14.6050,NanoLett.16.3533,J.Phys.D.AppliedPhysics.54.225304,PhysRevB.90.165403,APL.111.152402,PSSB.252.2395,arXiv.2105.06277}). Figs.~\ref{fig3}a and \ref{fig3}b demonstrate how a change in the RC time constant of the V$_+$ electrode has a direct impact on the spin-valve signal in both the X- and Y-channel. For this experiment $\alpha$ was not adjusted to create a virtual ground in the non-local part of the device (see explanation in section~\ref{CommonModeVoltageAndVirtualGround}). Instead, $\alpha$ was arbitrarily set to 1/2 which resulted in a non-vanishing common-mode voltage $V_\text{cm}(t)$ in this device (Fig.\,\ref{fig2}b).

\begin{figure*}[tb]
	\includegraphics[width=\linewidth]{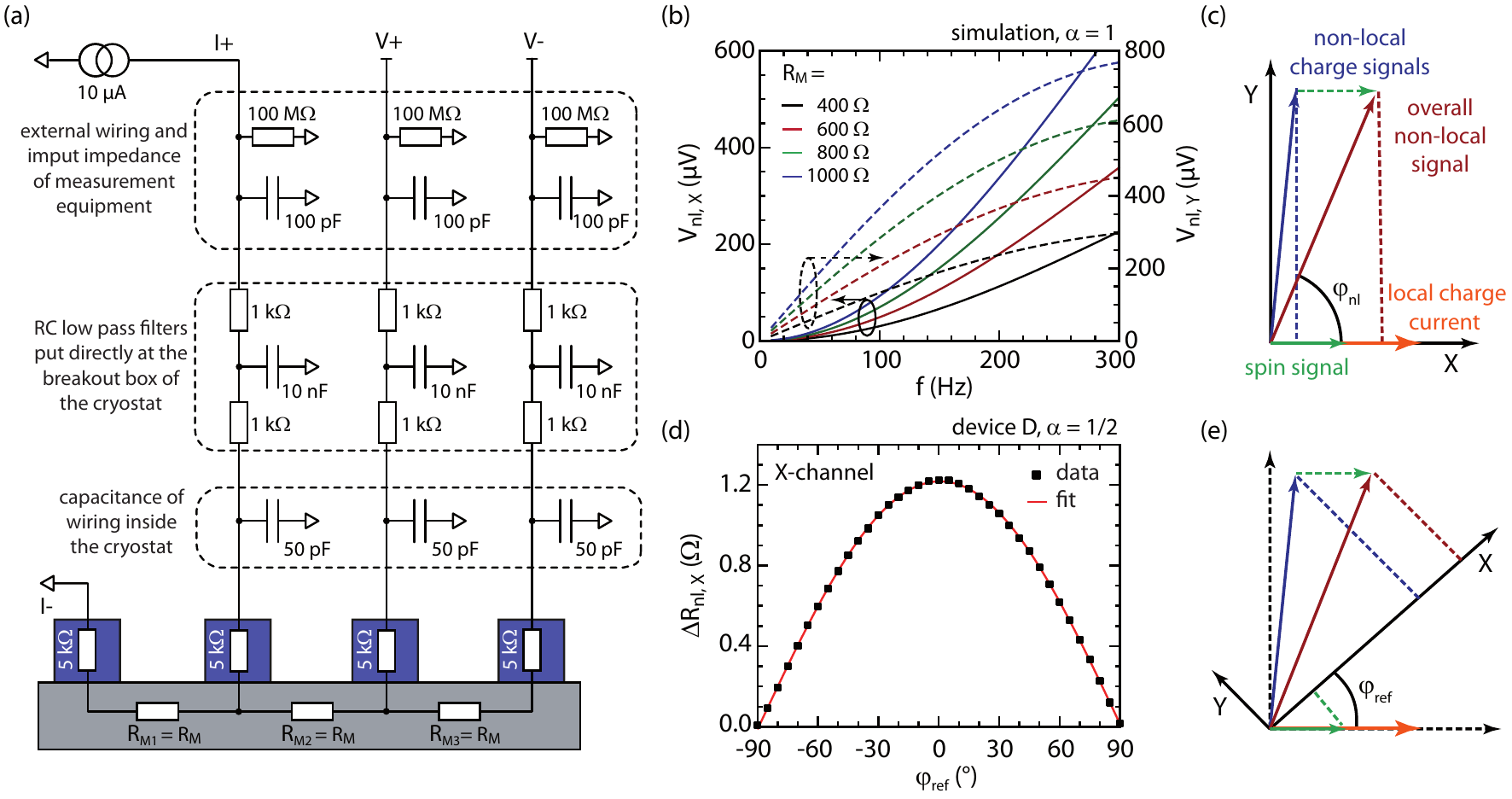}
	\caption{(a) Equivalent circuit that is used to simulate the detrimental impact of additional RC low pass filters on the non-local spin measurement. The high capacitances to ground of these filters significantly increase the spurious non-local signals that are caused by the charging and discharging currents over these capacitors. (b) The simulated signals in the X-channel (solid lines) and Y-channel (dashed lines) as a function of the applied measurement frequency for different resistances $R_\text{M}$ of the material. (c) and (e): To verify that the readout of the X-channel actually captures the whole spin signal, a rotation of the coordinate system by a reference phase $\varphi_{\text{ref}}$ can be utilized. As only the projection of each signal on the respective axis is measured, the amplitude of the spin signal should decrease as soon as the X-axis is rotated. (d) Corresponding experiment in which the amplitude of the spin-related switching in a spin valve measurement is recorded as a function of $\varphi_{\text{ref}}$. The maximum around 0$^\circ$ demonstrates that the presented current source creates a current signal and, therefore, a spin signal that is in phase with the output of the lock-in (compare to Fig.\,\ref{fig1}a).}
	\label{fig4}
\end{figure*}

The manipulation of the RC time constant was accomplished both by connecting ohmic resistors in series to the contact resistance of the V$_+$ electrode (resistance values in Figs.~\ref{fig3}a and \ref{fig3}b) and by increasing the length of the coaxial cable that is connecting the electrode to the differential voltage amplifier (\unit[100]{pF} per meter). These manipulations do not have any impact on the amplitude of the spin signal (switching in Fig.\,\ref{fig3}a indicated by the dashed arrows), but create an ever increasing offset in the signal of the X-channel. As the charge signal is shifted in phase by almost 90$^\circ$, the increase in the Y-channel signal (Fig.\,\ref{fig3}b) is significantly more pronounced. In fact, after applying a differential gain of $G_\textrm{diff}=1000$ by the voltage amplifier (see Fig.\,\ref{fig1}a), the voltage measured by the Y-channel of the lock-in eventually reaches \unit[1]{V}, which is the maximum voltage before the input of the used lock-in goes into an overload condition.

Simulations conducted with LTspice \cite{Zenodo} confirm the dependence of the non-local charge signal on variations in the resistances of the contacts and the capacitances of the wiring. The equivalent circuit for this simulation is shown in Fig.\,\ref{fig3}c, where either the contact resistance $R_\text{C3}$ or the capacitance $C_\text{W}$ of the wiring is varied, whereas all other values are set to parameters typical for a graphene-based spin-valve device (LTspice files can be found in Ref.~\cite{Zenodo}). From the simulations the amplitude $M=\sqrt{X^2+Y^2}$ and the phase $\varphi_\text{nl}$ (see definitions in Fig.\,\ref{fig1}c) of the differential, non-local voltage $V_\text{nl}=V_+ - V_-$ is extracted as a function of the driving frequency $f$ of the current in the injection circuit. Fig.\,\ref{fig3}d shows the results for different values of $R_\text{C3}$ and a fixed capacitance of $C_\text{W}=\unit[300]{pF}$, while Fig.\,\ref{fig3}e depicts the case with a fixed contact resistance $R_\text{C3}=\unit[10]{k\Omega}$ but varying capacitances $C_\text{W}$. Both cases demonstrate the linear increase in the amplitude of the non-local charge signal with the measurement frequency (see explanation in section~\ref{CapacitorCharging}). Furthermore, the simulations demonstrate that the resulting non-local charge signal has a phase close to but not exactly at 90$^\circ$ due to the ohmic contributions. The strongly decreasing phase towards lower frequencies is caused by the frequency-independent leakage currents that are flowing over the finite input impedances which are modelled by the $\unit[100]{M\Omega}$ resistors (see section~\ref{LeakageCurrents}). As soon as the signal that is caused by the capacitor charging becomes negligible towards $f=\unit[0]{Hz}$, the leakage currents, which are in phase with the current in the injection circuit, eventually are dominating and are pushing the phase to zero.

\begin{figure*}[tb]
	\includegraphics[width=\linewidth]{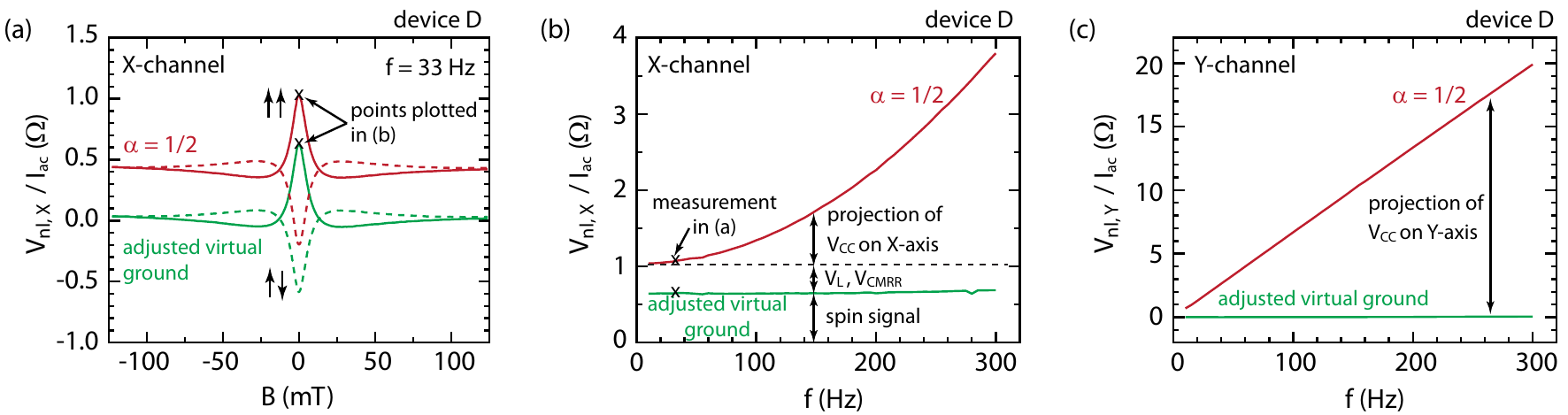}
	\caption{Demonstration that charge-induced signals can be significantly reduced by adjusting the virtual ground with the help of the presented current source. (a) Hanle curves measured with a symmetric bias applied to electrodes I$_+$ and I$_-$ (red curves) show a distinct background signal. Instead, the green curves depict the same Hanle measurement after the non-local part of the device was put to a virtual ground. This prevents the flow of any charge current in the detection circuit over resistances or capacitances to ground and, therefore, diminishes the charge-induced background signals. (b) and (c): Non-local signal as a function of measurement frequency for both the X-channel (b) and the Y-channel (c). For these graphs the values in the parallel configuration of the Hanle curves at $B=\unit[0]{T}$ were measured for both bias conditions. In case of a symmetric bias voltage, the charge-induced signal gets much larger than the actual spin signal, especially in the Y-channel. Instead, in case of the adjusted virtual ground, the signal in the Y-channel is negligible over the whole measurement range and the signal in the X-channel consists only of the actual spin signal.}
	\label{fig5}
\end{figure*}

The experimental results and simulations shown in Fig.\,\ref{fig3} demonstrate how much the measurement signal can depend on the wiring of the measurement setup. In particular, RC low pass filters can be extremely disadvantageous to non-local, lock-in-based measurements, as they significantly increase the total capacitance of the V$_+$ and V$_-$ lines to GND. This in turn increases the spurious signals that are caused by the charging currents of the corresponding capacitors. We use the equivalent circuit that is shown in Fig.\,\ref{fig4}a to simulate how a mere change of the DUT's resistances ($R_\text{Mi} = R_\text{M}$, for $i=1,2,3$) impacts the non-local signal. In this equivalent circuit the charge signal was intentionally diminished by assuming identical values for contact resistances, capacitances of the wiring, and RC filter characteristics for all relevant contacts. Therefore, the only difference in the overall RC-time-constants of contacts V$_+$ and V$_-$ is given by the value of $R_\text{M3}=R_\text{M}$. Of course, the resistance $R_\text{M1}=R_\text{M}$ in the injection circuit also influences the common-mode voltage $V_\text{cm}$ which is driving the non-local currents in the first place. In Fig.\,\ref{fig4}b the resulting non-local charge signals for both the X-channel (solid lines) and Y-channel (dashed lines) are plotted as a function of the measurement frequency for different values of $R_\text{M}$. The results strikingly demonstrate how significantly the spurious non-local charge signal depends in this case even on gate-induced changes in the DUT's resistivity. The simulations were conducted with an applied current of $I=\unit[10]{\mu A}$, therefore the left y-axis in Fig.\,\ref{fig4}b corresponds to a non-local resistance between 0 and $\unit[60]{\Omega}$.

The example of the RC filters illustrates how drastically specific details of the setup can influence such measurements. We therefore strongly advice to verify the basic assumptions that are often made for individual setups. One important assumption is that the spin signal is fully located within the X-channel of the lock-in. As discussed in section~\ref{Principal-of-a-non-local-spin-measurement}, the spin signal should be in phase with the current in the injection circuit, which in turn should be in phase with the output signal from the lock-in's internal oscillator (green and orange vectors in Fig.\,\ref{fig4}c). Instead the vector addition of all non-local charge signals results in a signal with a phase somewhere between 0$^\circ$ and 90$^\circ$ (blue vector in Fig.\,\ref{fig4}c).

Most lock-in amplifiers offer the possibility to rotate the X-Y-coordinate system by a reference phase $\varphi_{\text{ref}}$ relative to the internal oscillator, which is illustrated in Fig.\,\ref{fig4}e. The measured signals are then the projections of both the spin and charge signals onto the respective axis. The amplitude of the actual spin signal in the X-channel as a function of $\varphi_{\text{ref}}$ can be determined by the amplitude of the switching in a spin-valve measurement that is repeated for different $\varphi_{\text{ref}}$ \cite{2DMaterials.2.024001}. In case of our setup and the newly designed current source, the result of such an experiment is shown in Fig.\,\ref{fig4}d. The actual data points (black squares) can be perfectly fitted (red line) with an equation that describes the projection of the spin signal onto the rotating X-axis:
\begin{equation}
    \Delta R_\text{nl,X}(\varphi_\text{ref}) = \Delta R_\text{nl,X}^0\cdot\cos\left( \varphi_\text{ref}-\varphi_\text{off} \right),
\end{equation}
with a maximum spin signal $\Delta R_\text{nl,X}^0$ and an offset phase $\varphi_\text{off}$ between the spin signal and the internal oscillator signal. The fit yields a value of $\varphi_\text{off}=\unit[0.6]{^\circ}$, which demonstrates that our current source creates a current signal and, therefore, a spin signal that is almost perfectly in phase with the output of the lock-in.

Finally, we demonstrate the actual strength of such a current source: The distribution of the bias voltage between contacts I$_+$ and I$_-$ can be controlled in a way, that the non-local part of the device is set to a virtual ground (see theoretical explanation in section~\ref{CommonModeVoltageAndVirtualGround} and discussion about the technical operation of the current source in the Supplemental Material \cite{Supplement}). Under this condition the common-mode voltage in the non-local part is zero at all times ($V_\text{cm}(t)=0$), i.e. no charge current can flow over any resistance or capacitance to GND. Furthermore, spurious signals due to a finite common-mode rejection ratio of the measurement setup are also minimized.

The impact of such an adjustment procedure on a Hanle measurement is shown in Fig.\,\ref{fig5}a. The particular device for this demonstration was chosen because it only exhibits a constant background signal, which is typical for the capacitor charging effect. The absence of a magnetic field dependent background indicates that the current spreading effect is largely diminished. The red lines depict Hanle curves in both parallel (solid lines) and anti-parallel (dashed lines) magnetization configurations of the I$_+$ and V$_+$ electrodes if a symmetric bias is applied to electrodes I$_+$ and I$_-$ ($\alpha = 1/2$, i.e. $1/2\cdot V_\text{bias}(t)$ to I$_+$ and $-1/2\cdot V_\text{bias}(t)$ to I$_-$, compare to Fig.\,\ref{fig2}b). As the growth of uniform and reproducible tunnel barriers on the inert graphene is quite challenging \cite{ScientificReports.5.14332,AdvancedMaterialsInterfaces.4.1700232,JApplPhysics.117.083909,ACSNano.8.7704,APLMaterials.4.046104}, the difference in the respective contact resistances of I$_+$ and I$_-$ is usually quite large. Therefore, the virtual ground point in case of a symmetric bias voltage normally does not lie within the transport channel of the DUT. Accordingly, there is a finite common-mode voltage $V_\text{cm}(t)$ driving the spurious non-local signals, which is seen as a background signal. Instead, the green curves in Fig.\,\ref{fig5}a depict the same Hanle measurement (recorded in the X-channel) after $\alpha$ was adjusted in such a way that the signal in the Y-channel of the lock-in was minimized. With this adjustment also the background signal in the Hanle measurement has almost completely vanished (in the Supplemental Material \cite{Supplement} we demonstrate that adjusting $\alpha$ until the signal in the Y-channel is minimized also avoids common-mode related artifacts in local measurements).

The dependence of the non-local signal on the frequency is depicted in Figs.~\ref{fig5}b and \ref{fig5}c for the X-channel and Y-channel, respectively. For these graphs the values in the parallel configuration of the Hanle curves at $B=\unit[0]{T}$ were measured for both the symmetric (red) and the adjusted bias condition (green). In the Y-channel (Fig.\,\ref{fig5}c) the aforementioned linear increase of the charge-induced, non-local signal with increasing frequency can be observed in case of the symmetric bias condition. At a frequency of $f=\unit[300]{Hz}$ the charge-induced signal is already over 30x larger than the actual spin signal in the X-channel. For more unfavourable values of $\alpha$ or in devices that have even more strongly varying contact resistances, the charge-induced signal can even be much larger. This and the fact that spin signals can be much smaller than the one presented in Fig.\,\ref{fig5} can yield huge differences between the amplitudes of the spin and charge-induced signals of up to several orders of magnitude.

It is important to note that the charge-induced signal in Fig.\,\ref{fig5}c goes to zero for $f\rightarrow \unit[0]{Hz}$ even in case of $\alpha=1/2$. This implies that the complete signal in the Y-channel is caused by the capacitor charging effect, as its amplitude scales linearly with the applied frequency (see explanation in section~\ref{CapacitorCharging}). On the other hand, in Fig.\,\ref{fig5}b there is still an offset between the two curves for $\alpha=1/2$ and the adjusted ground condition for $f\rightarrow \unit[0]{Hz}$ (indicated by the dashed line). That means that there are additional charge-induced signals present in the X-channel, apart from the projection of $V_\text{CC}$ onto the X-channel that is responsible for the increasing background signal towards higher frequencies. This increase is not linear as the one seen in Fig.\,\ref{fig5}c, because the overall phase of $V_\text{CC}$ varies slightly with the applied frequency (see frequency-dependent phases in Figs.~\ref{fig3}d and \ref{fig3}e, and the non-linear increase of the signal in the X-channel in Fig.\,\ref{fig4}b).
Importantly, adjusting the virtual ground in such a way that the spurious signal in the Y-channel vanishes, also completely removes these additional signals within the X-channel, i.e. the green curve in Fig.\,\ref{fig5}b only consists of the actual spin signal. Therefore, the offset in case of $\alpha=1/2$ for $f\rightarrow \unit[0]{Hz}$ must be caused by the other charge-induced artifacts that depend on the common-mode voltage, namely $V_\text{L}$ and $V_\text{CMRR}$ (see sections~\ref{LeakageCurrents} and \ref{CommonModeRejectionRatio}). These two signals can appear not only in AC but also DC ($f\rightarrow \unit[0]{Hz}$) measurements (see Supplemental Material \cite{Supplement} for a more detailed discussion). The possibility to remove these spurious signals therefore demonstrates that the presented current source is also highly beneficial for DC measurements.

\section{Conclusion}
\label{Conclusion}
We have discussed seven different charge-induced, non-local voltage signals that can appear in a spin measurement, despite the fact that this measurement scheme is repeatedly claimed to be free of such spurious signals. However, it is important to emphasize that the relevance of each individual charge-induced signal can vary significantly depending on details in device fabrication, properties of the investigated material, device geometry, measurement setup, and measurement technique. In our study, we focused on experimental data that was measured with a lock-in based technique on graphene spin-valve devices. In such devices we observe that two charge-induced signals normally prevail: the signal that is caused by current-spreading (section~\ref{CurrentSpreading}) and the signal that is caused by the capacitor charging currents (section~\ref{CapacitorCharging}). These two signals dominate because tunnel barriers must be used in these devices to overcome the conductivity mismatch problem for an efficient spin injection \cite{PhysRevB.62.R16267,PhysRevB.62.R4790,PhysRevB.64.184420}. But it is known that growing homogeneous, pinhole-free tunnel barriers on top of the inert graphene surface is quite challenging without damaging or modifying the underlying graphene  \cite{ScientificReports.5.14332,AdvancedMaterialsInterfaces.4.1700232,JApplPhysics.117.083909,ACSNano.8.7704,APLMaterials.4.046104}. Pinholes in the tunnel barrier promote the effect of current spreading, whereas strongly varying contact resistances lead to different RC time constants and, therefore, the signal that is caused by the capacitor charging currents. Furthermore, when tunnel barriers are present, a larger bias voltage must be applied to drive the same current as without barriers. Without the possibility to adjust the virtual ground, this unavoidably also increases the common-mode voltage $V_\text{cm}(t)$ (Fig.\,\ref{fig2}b) and, therefore, all non-local charge signals that are driven by $V_\text{cm}(t)$. In devices without tunnel barriers the relative contribution of each individual charge-induced signal may be completely different.

Nevertheless, the occurrence of all these different charge-induced signals is a general problem, as in the past these artifacts may have been falsely attributed to a variety of spin- or valley-related transport phenomena. For example, it has already been reported that measurement artifacts have most likely led to erroneous and misleading claims regarding the value of the spin Hall angle in certain materials or the spin transport properties of topological insulators. Therefore, understanding both the origins of these signals and the specific conditions under which they can appear is crucial for the correct analysis of non-local transport experiments, not only in the field of spintronics but also other newly emerging fields as valleytronics. If available, we have discussed ways how these spurious signals can be minimized. In particular, we demonstrated that our current source, that is able to create a virtual ground in the non-local part of the device, can remove any charge-induced signals that are caused by a common-mode voltage.\\ \\

\begin{acknowledgments}
This project has received funding from the European Union's Horizon 2020 research and innovation programme under grant agreement No. 881603 (Graphene Flagship) and the Deutsche Forschungsgemeinschaft (DFG, German Research Foundation) under Germany's Excellence Strategy - Cluster of Excellence Matter and Light for Quantum Computing (ML4Q) EXC 2004/1 - 390534769, through DFG (BE 2441/9-1 and STA 1146/11-1), and by the Helmholtz Nano Facility (HNF) \cite{HNF} at the Forschungszentrum J\"ulich.
\end{acknowledgments}

\end{document}


\title{Supplemental Material:\\Charge-induced artifacts in non-local spin transport measurements:\\How to prevent spurious voltage signals}

\author{Frank Volmer}
\affiliation{2nd Institute of Physics and JARA-FIT, RWTH Aachen University, 52074 Aachen, Germany}
\affiliation{AMO GmbH, Advanced Microelectronic Center Aachen (AMICA), 52074 Aachen, Germany}

\author{Timo Bisswanger}
\affiliation{2nd Institute of Physics and JARA-FIT, RWTH Aachen University, 52074 Aachen, Germany}

\author{Anne Schmidt}
\affiliation{2nd Institute of Physics and JARA-FIT, RWTH Aachen University, 52074 Aachen, Germany}

\author{Christoph Stampfer}
\affiliation{2nd Institute of Physics and JARA-FIT, RWTH Aachen University, 52074 Aachen, Germany}
\affiliation{Peter Gr\"unberg Institute (PGI-9), Forschungszentrum J\"ulich, 52425 J\"ulich, Germany}

\author{Bernd Beschoten}
\affiliation{2nd Institute of Physics and JARA-FIT, RWTH Aachen University, 52074 Aachen, Germany}

\maketitle

\begin{figure*}[tb]
	\includegraphics[width=\linewidth]{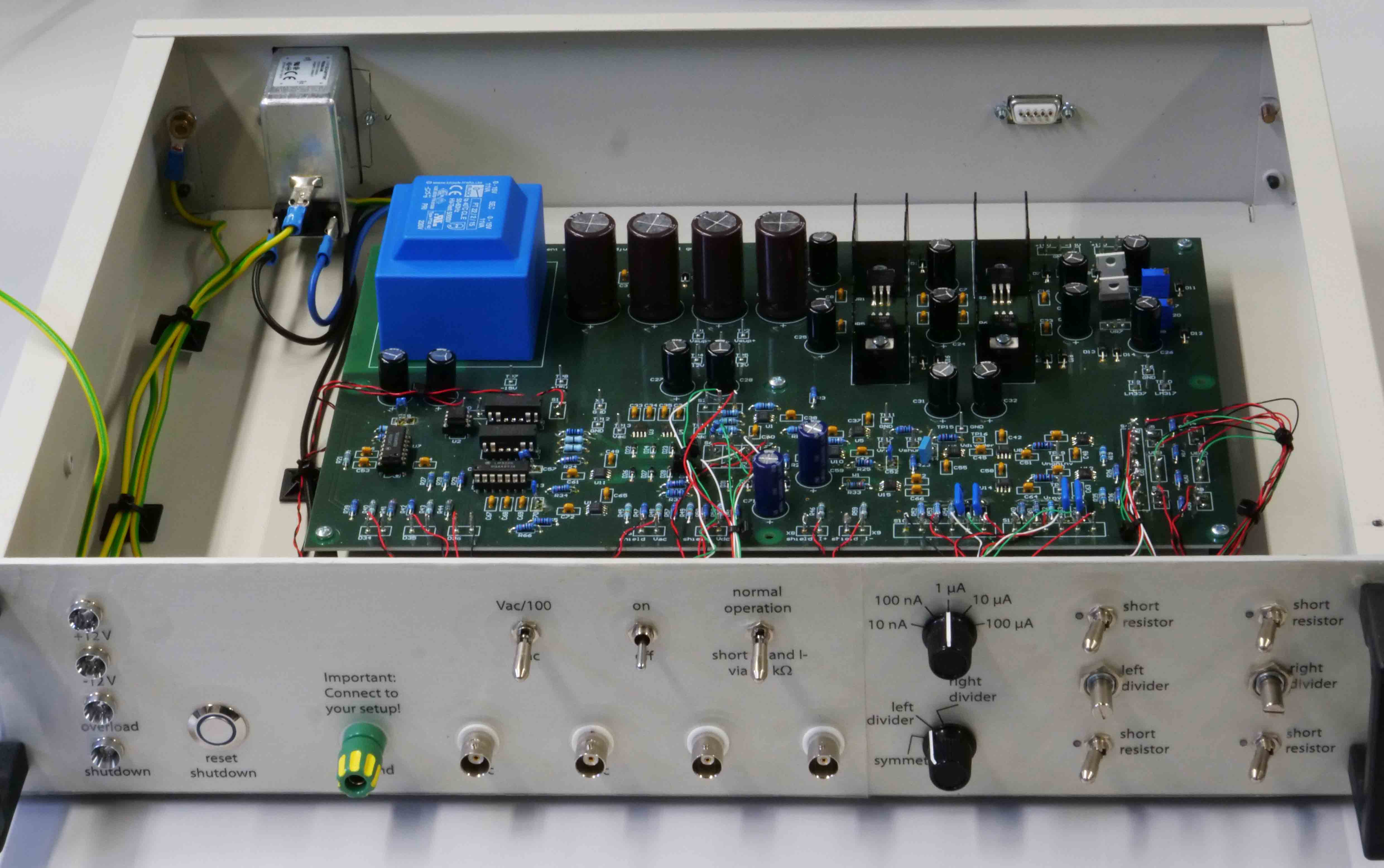}
	\caption{Potograph of the assembled current source.}
	\label{FigS1}
\end{figure*}

\begin{figure*}[tb]
	\includegraphics[width=\linewidth]{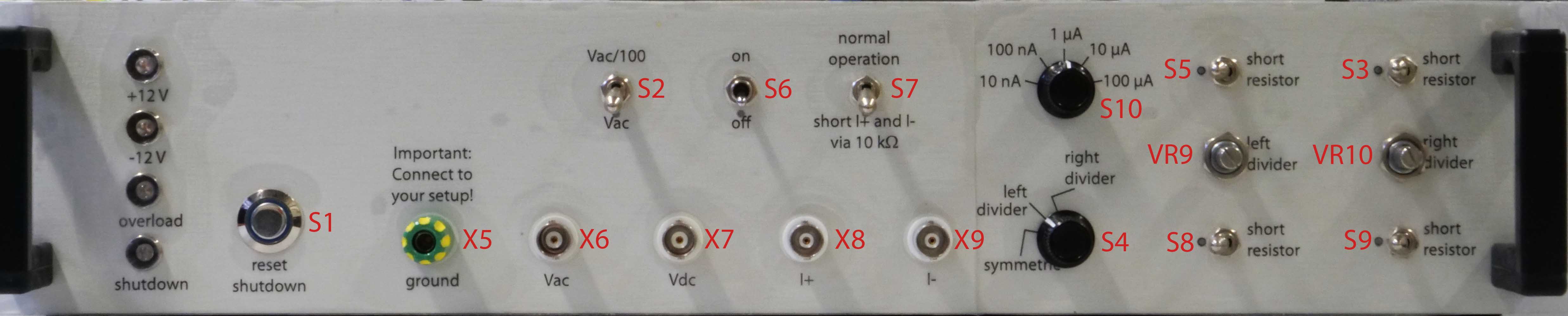}
	\caption{Front panel of the current source with reference designators used in the schematics in Figs.~\ref{FigS9} to \ref{FigS14}.}
	\label{FigS2}
\end{figure*}

In this Supplemental Material, we first discuss the basic operation of our voltage-controlled current source in sections \ref{PowerSupply} to \ref{CurrentSource} (pictures of the assembled electronic are shown in Figs.~\ref{FigS1} and \ref{FigS2}). Then, we present measurements about its accuracy (section~\ref{DCAccuracy}), bandwidth (section~\ref{ACperformance}), stability during fast transients (section \ref{TurnOnOff}), and over-voltage protection (section~\ref{OVP}). In section \ref{VcmLocalMeasurement}, we discuss how a common-mode voltage can even impact local measurements before we give practical hints on how to adjust the virtual ground in section~\ref{Procedure-to-adjust-the-virtual-ground}. After discussing the applications and limitations of the current source in section~\ref{Applications-Limitations}, we finally give some practical advices both on the use of the current source and on the investigation of charge-induced signals in non-local measurements in section~\ref{FinalNotes}.

All schematics necessary to rebuilt the current source can be found in Figs.~\ref{FigS9} to \ref{FigS14} at the end of this Supplemental Material and will be discussed in the following sections. On Zenodo (\url{https://doi.org/10.5281/zenodo.6525020}) we further provide: 1.) Gerber files of the PCB used for our current source, 2.) the corresponding Altium Designer files, 3.) a bill of materials (BOM), 4.) the LTspice models for each simulation, and 5.) the data used to create each figure both in the main manuscript and this Supplemental Material.


\section{Power supply}
\label{PowerSupply}
A traditional linear power supply that is based on a safety isolating transformer and linear voltage regulators is used to power the current source (see Fig.\,\ref{FigS9}). The series connection of the linear voltage regulators is cascading the ripple rejection of each regulator. Although this is a quite energy-inefficient design, it nevertheless provides an easy way to achieve voltage rails that have a minimum amount of both ripples and high frequency noise. The first stage ($V_\text{dc}=\unit[\pm 15]{V}$) is used to power reed relays in an over-voltage protection circuit, the second stage ($V_\text{dc}=\unit[\pm 12]{V}$) is used to power the operational and instrumentation amplifiers, and the final third stage (variable output voltage) is used as a low-noise voltage rail connected to the output of the current source via low-leakage clamping diodes (detailed explanation in the following sections).

We note that the power supply depicted in Fig.\,\ref{FigS9} is able to provide a larger current than is actually needed for the operation of the current source, as the power supply is also intended to supply additional equipment. If only the current source is to be reproduced, we recommend to use other types of linear regulators with lower output currents but better characteristics with respect to output voltage noise and power supply rejection ratio. Also, the supply voltage for the operational and instrumentation amplifiers may be reduced depending on the envisioned application. The minimum supply voltages $\pm V_\text{sup}$ for the amplifiers, at which a reliable operation of the current source is guaranteed regardless of the position of the virtual ground, is given by:
\begin{equation}
    \pm V_\text{sup} = \pm \left( 2\cdot \left|V_\text{ref,max}\right| + \left|V_\text{DUT,max}\right| + \unit[2]{V} \right) ,
\end{equation}
with $V_\text{ref,max}$ being the maximum of the externally applied reference voltage $V_\text{ref}$ which will be discussed in section~\ref{ReferenceVoltages} and $V_\text{DUT,max}$ being the maximum voltage drop across the device under test (DUT). The additional voltage of $\unit[2]{V}$ is providing headroom for the amplifiers (see input voltage range and output voltage swings in the respective data sheets).

To prevent ground loops, we do not directly connect the protective earth (PE) of the mains to the signal ground (GND) inside the electronic. Nevertheless, it is of upmost importance to have a low-impedance connection between the GND of the current source and the common ground of the measuring setup (connector X5 in Fig.\,\ref{FigS9} can be used for this connection). We highly advise to use a star ground layout for this purpose. Neglecting such a low-impedance connection can lead to significant instabilities, especially for smaller output currents in the nA-range.

\section{Reference voltages}
\label{ReferenceVoltages}

One important requirement for the current source is that the time-dependent current $I_\text{DUT}(t)$ through the DUT has to be controlled by an externally applied reference voltage $V_\text{ref}(t)$. As will be explained later on, the connection between $I_\text{DUT}(t)$ and $V_\text{ref}(t)$ is given as $I_\text{DUT}(t) = V_\text{ref}(t) / R_\text{shunt}$, with varying shunt resistors $R_\text{shunt}$ for different current levels.

As the current source is intended to be used in conjunction with a lock-in amplifier, the typical sinusoidal output voltage $V_\text{ac}(t)$ of a lock-in amplifier is one part of this reference voltage. To also enable bias dependent measurements, a DC voltage $V_\text{dc}$ has to be superimposed to this sinusoidal signal. However, the excitation amplitude of the AC signal has to be much smaller than the one of the DC signal to guarantee that the AC signal does not impact the bias dependent measurement too strongly. Instead of using small amplitudes of the external AC voltage, it is desirable to attenuate the external AC signal within the electronic. This procedure will also attenuate every interference which is coupled into the AC signal on the way from the lock-in amplifier into the current source. Hence, the overall reference voltage is designed to be:
\begin{equation}
    V_\text{ref}(t)=\beta \cdot V_\text{ac}(t)+V_\text{dc},
\end{equation}
with $\beta$ being either 1 or 0.01. This is accomplished by the circuit shown in Fig.\,\ref{FigS10}, which will be discussed now in detail.

In order to decouple the external voltage sources from the current source electronic (e.g.\,to prevent ground loops) instrumentation amplifiers are used (IC3 and IC4). These ICs measure the differential voltage between shield and inner conductor of the BNC connectors (X6 and X7) and make a conversion to a single-ended voltage signal which is referenced to the signal ground of the current source electronic. Nevertheless, we recommend that the external voltage sources which are providing the voltages $V_\text{ac}(t)$ and $V_\text{dc}$ are not floating but referenced to the same ground potential as the rest of the measurement system. Therefore, both the shields of the BNC connectors X6 and X7 and the low-side output of the voltage source should be put to the common ground node of the measuring system via a star ground layout. By using coaxial cables, this configuration provides good immunity to capacitively coupled interference signals.

In case that one of the two voltage inputs ($V_\text{ac}$ or $V_\text{dc}$) is not used, $\unit[10]{M\Omega}$ resistors (R45, R48, R49, R58) prevent this input from being floating and tie it to signal ground. This makes the input more robust against externally coupled interference signals. Nevertheless, it is good practice to additionally use shorting BNC end caps to shorten the inner conductor and the shield of the BNC connector, pushing any differential voltage measured by the instrumentation amplifier to zero. This practice is especially important if the $\unit[10]{M\Omega}$ resistors may be replaced with higher value ones in order to decouple the external voltage sources even further from the signal ground of the current source.

\begin{figure*}[htb]
	\centering
	\includegraphics[width=1\textwidth]{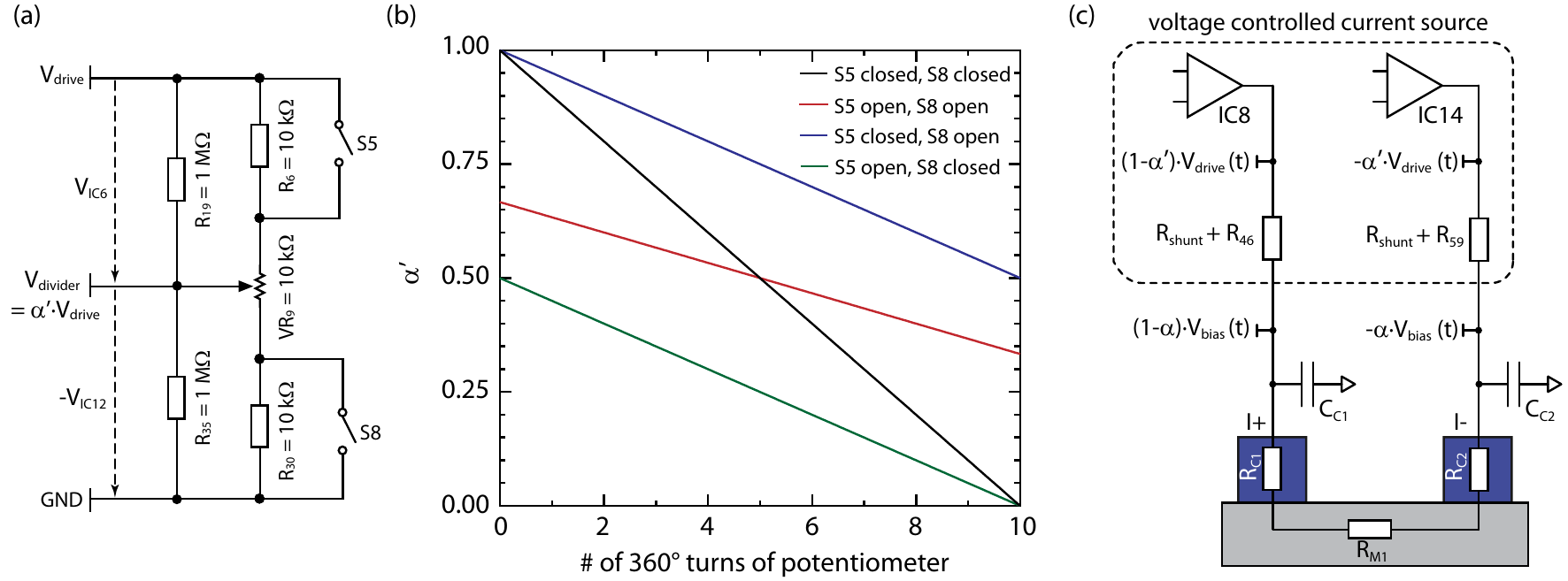}
	\caption{a) Schematic of one of the manually adjustable voltage dividers, combining the most crucial components from the schematics in Figs.~\ref{FigS11} and \ref{FigS12}. Depending on both the wiper position of VR9 and the positions of S5 and S8, the voltage $V_\text{divider} = \alpha' \cdot V_\text{drive}$ at the wiper can be set to be anywhere between $V_\text{drive}$ ($\alpha'=1$) and ground potential, i.e.\,\unit[0]{V} ($\alpha'=0$). b) Corresponding values of $\alpha'$ if a 10 turn potentiometer is used for VR9 for different switch positions of S5 and S8. c) Connection between $\alpha'$ and $\alpha$ of the main manuscript: As long as a current is flowing, the output voltages of IC8 and IC14, which can be freely adjusted via $\alpha'$, differ from the applied voltages to the outputs of I$_+$ and I$_-$ by the voltage drop over the shunt resistors and the $\unit[100]{\Omega}$ protection resistors R46 and R59 (compare to Fig.\,\ref{FigS11}).}
	\label{FigS3}
\end{figure*}

The mixing of the AC and DC voltage signals is accomplished by an inverting summing amplifier circuit, consisting of IC1 with resistors R4, R5, R7 and R9. Switching between R5 and R7 via switch S2 is setting the attenuation factor $\beta$ to either 1 or 0.01.

To make sure that the current signal is in phase with the external voltage signal and not inverted by $\unit[180]{^\circ}$, an inverting amplifier with gain 1 is used afterwards (IC10 with R10 and R16). Finally, a simple RC low-pass filter (R25, C52) is used for rudimentary filtering of any voltage spikes or high frequency interference signals.

\section{Current source: Principle operation}
\label{CurrentSource}
The circuit diagram of the actual current source is depicted in Fig.\,\ref{FigS11}. The reference voltage $V_\text{ref}(t)=\beta \cdot V_\text{ac}(t)+V_\text{dc}$, that was described in the last section, is put to the non-inverting input of IC9 via node "Vref". The voltage applied to the inverting input of IC9 is the voltage drop over a shunt resistor measured by the instrumentation amplifier IC15 with a gain of 1 (set by an open connection between pins 2 and 3 of IC15). The value of the shunt resistor is set by the rotary switch S10.

The output of IC9 now tries to minimize the voltage difference between its inverting and non-inverting input. This can be achieved if the voltage drop over the shunt resistor is equal to $V_\text{ref}(t)$. As the shunt resistor is put in series to the DUT (which will be connected between the BNC connectors X8 and X9), the desired relation
\begin{eqnarray}
    I_\text{DUT}(t) &=& V_\text{ref}(t) / R_\text{shunt} \\ \nonumber
    &=& \left[ \beta \cdot V_\text{ac}(t)+V_\text{dc} \right] / R_\text{shunt}
\end{eqnarray}
of the voltage controlled current source is achieved.

The current $I_\text{DUT}(t)$ itself is sourced and drained by the outputs of the operational amplifiers IC8 and IC14, which are wired as voltage followers. As it becomes apparent further below, the output voltage of IC9 referenced to GND, $V_\text{drive}(t)$, is the same voltage that is needed for driving the current through every resistance between the outputs of IC8 and IC14.

The output voltage of IC9 is applied to a voltage divider referenced to GND. In the schematic in Fig.\,\ref{FigS11} only one part of the overall voltage divider is shown (the $\unit[1]{M\Omega}$ resistors R19 and R35), whereas the manually adjustable part of the divider is depicted in Fig.\,\ref{FigS12}. The current source is equipped with three different sets of ohmic voltage dividers which can be selected by switch S4 in Fig.\,\ref{FigS12}. In the following we discuss the situation in which S4 is set to position 2.

The corresponding schematic is depicted in Fig.\,\ref{FigS3}a. The two $\unit[1]{M\Omega}$ resistors (R19 and R35 in Fig.\,\ref{FigS11}) are solely a backup in case that the wiper of the potentiometer or the switch S4 (not included in Fig.\,\ref{FigS3}a) may fail. In such a case, the $\unit[1]{M\Omega}$ resistors guarantee that the output of IC9 ($V_\text{drive}$) has a path to ground and that the voltage $V_\text{divider}$ is never floating.

Instead, under normal operation the lower value resistors R6, VR9 and R30, which are put in parallel to either R19 or R35 (compare Figs.~\ref{FigS11} and \ref{FigS12} to \ref{FigS3}a), determine the characteristic of the overall voltage divider. Both the position of the potentiometer and the positions of the switches S5 and S8 determine the voltage at node $V_\text{divider}$ as $V_\text{divider}= \alpha'\cdot V_\text{drive}$ with $0 \leq \alpha' \leq 1$.

For example, if S5 and S8 are closed and, therefore, short R6 and R30, the voltage $V_\text{divider}$ can be set to be anywhere between $V_\text{drive}$ ($\alpha'=1$) and ground potential, i.e.\,\unit[0]{V} ($\alpha'=0$), depending on the position of the wiper. As we use a 10 turn potentiometer (wirewound with linear characteristic), the resulting value of $\alpha'$ as a function of the wiper position follows the black line in Fig.\,\ref{FigS3}b. Opening one or both of the switches S5 and S8 limit the range over which $\alpha'$ can be tuned, therefore increasing the sensitivity at which $\alpha'$ can be set by turning the potentiometer by a certain angle (see corresponding curves in Fig.\,\ref{FigS3}b).

The respective voltage differences between nodes $V_\text{drive}$ and $V_\text{divider}$ on the one hand and $V_\text{divider}$ and GND on the other hand are then measured via the precision instrumentation amplifiers IC6 and IC12 in Fig.\,\ref{FigS11} with a gain of one. It is important that the inputs of IC6 and IC12 are switched in respect to each other (both inverting inputs are connected to node $V_\text{divider}$). Therefore, there is a sign reversal between the voltage that is measured by IC12 and the voltage $V_\text{drive}$ that is applied to the voltage divider.

The voltage $V_\text{drive}(t)$ is therefore divided into two components:
\begin{eqnarray}
 V_\text{drive}(t) &=& V_\text{IC6}(t) - V_\text{IC12}(t) \\ \nonumber
 &=& \left[ (1-\alpha')\cdot V_\text{drive}(t)\right] - \left[ -\alpha'\cdot V_\text{drive}(t) \right],
\label{eq:V_drive}
\end{eqnarray}
with a voltage divider ratio $0 \leq \alpha' \leq 1$. The two voltages $V_\text{IC6}(t)$ and $V_\text{IC12}(t)$ are then buffered (see operational amplifiers IC8 and IC14 at the outputs of the instrumentation amplifiers IC6 and IC12 in Fig.\,\ref{FigS11}). The operational amplifiers IC8 and IC14 are accordingly sourcing and draining the current $I_\text{DUT}(t)$ through the DUT.

But as long as a current is flowing, the two voltages $V_\text{IC6}(t)$ and $V_\text{IC12}(t)$ are not the ones which are applied to the outputs $I_+$ and $I_-$ of the current source. Under normal operation, also the voltage drop over the shunt resistors and the $\unit[100]{\Omega}$ protection resistors R46 and R59 has to be considered. This is depicted in the schematic in Fig.\,\ref{FigS3}c. Accordingly, the voltage divider ratio $\alpha'$ is not equal to the coefficient $\alpha$ discussed in the main manuscript. Nevertheless, changing $\alpha'$ will change $\alpha$ in a continuous way, too.

\section{DC accuracy}
\label{DCAccuracy}

The accuracy of the current source was measured by connecting a Keithley 2450 source meter unit in current measuring mode between the connections "I+" and "I-" of Fig.\,\ref{FigS11}, i.e.\,using the Keithley 2450 as the DUT. Then a Yokogawa 7651 voltage source was used to apply a varying voltage $V_\text{dc}$ for different shunt resistor's values. To achieve the best-possible accuracy even for lower values of $V_\text{dc}$, a shorting BNC end cap should be used for the unused BNC connector labeled by $V_\text{ac}$ in Fig.\,\ref{FigS10}. This end cap minimizes the voltage offset of IC3, which otherwise yields a systematic small error caused by the mixing of $V_\text{dc}$ and $V_\text{ac}$ to $V_\text{ref}$ (see section~\ref{ReferenceVoltages}).

\begin{figure}[tb]
	\centering
	\includegraphics[width=0.5\textwidth]{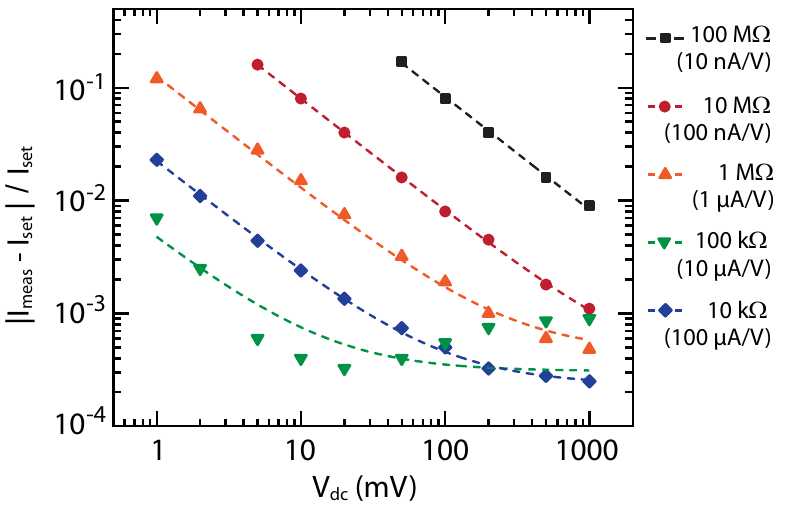}
	\caption{DC accuracy of the current source: The difference between the measured current $I_\text{meas}$ and the set current $I_\text{set} = V_\text{dc} / R_\text{shunt}$ is plotted as a relative deviation. Symbols represent measurements, whereas dashed lines are fits.}
	\label{FigS4}
\end{figure}

The difference between the actually measured current $I_\text{meas}$ and the set current $I_\text{set} = V_\text{dc} / R_\text{shunt}$ (with $R_\text{shunt}$ being the nominal shunt resistor value) is plotted in the double logarithmic plot of Fig.\,\ref{FigS4} as a relative deviation
\begin{equation}
    \frac{\Delta I}{I_\text{set}} = \frac{\left|I_\text{meas}-I_\text{set}\right|}{I_\text{set}}\;.
\end{equation}
The symbols represent measurements, whereas the dashed lines are fits assuming that the measured current can be expressed as
\begin{equation}
    I_\text{meas} = \left(1+\epsilon\right)\cdot I_\text{set} + I_\text{err}\;,
\end{equation}
with a scaling error $\epsilon$ and an error current $I_\text{err}$.

The scaling error $\epsilon$ gets noticeable in the flattening of the curves in Fig.\,\ref{FigS4} towards higher $V_\text{dc}$ values. The fits yield scaling errors in the range of $10^{-4}$ to $10^{-3}$ which is in complete accordance to the chosen tolerance class of the shunt resistors (real resistor values are within 0.1\% of the nominal values). In contrast, the error current $I_\text{err}$ gets noticeable in the increase of the relative deviation towards lower $V_\text{dc}$ values, as its relative contribution increases towards lower applied currents. For shunt resistor values between $\unit[1]{M\Omega}$ and $\unit[100]{M\Omega}$ the fits yield quite similar error currents in the range of $\unit[80]{pA}$ to $\unit[120]{pA}$. Interestingly, the fit does not follow the data at all for a shunt resistance of $\unit[100]{k\Omega}$ (green curve), before it is good again for $\unit[10]{k\Omega}$ (blue curve). Importantly, the fit to the $\unit[10]{k\Omega}$ shunt resistor yields a significantly increased error current of $\unit[2.2]{nA}$.

Both, the existence of two regimes of significantly different error currents and the odd behaviour of the curve in between, can be well understood by two important device characteristics of the instrumentation amplifier IC15, which measures the voltage drop over the shunt resistor in Fig.\,\ref{FigS11}. These two characteristics are the input offset current $I_\text{ioc}$ and the input offset voltage $V_\text{iov}$, respectively. As explained further below, the error current in most cases can be approximated by:
\begin{equation}
    I_\text{err} \approx \text{max}\left(  I_\text{ioc}\;, \; \frac{V_\text{iov}}{R_\text{shunt}} \right) \;.
\end{equation}

The input offset current $I_\text{ioc}$ is the difference in the currents that flow in or out of each of the two inputs of the instrumentation amplifier. This current is unavoidably drained into or sourced out of the circuit, which contains the DUT in Fig.\,\ref{FigS11}. Therefore, the lower the applied current over the DUT (i.e.\,the lower the reference voltage $V_\text{ref}= V_\text{dc}$ or the higher the shunt resistor value $R_\text{shunt}$ according to $I_\text{DUT}(t) = V_\text{ref}(t) / R_\text{shunt}$) the higher the error becomes that is caused by the input offset current of the instrumentation amplifier. The extracted error currents of around \unit[100]{pA} for shunt resistor values between $\unit[1]{M\Omega}$ and $\unit[100]{M\Omega}$ are accordingly in very good agreement to the data sheet's value of the input offset current of IC15.

On the other hand, the input offset voltage $V_\text{iov}$ is defined as the voltage that must be applied between the two inputs of the instrumentation amplifier to obtain a perfectly vanishing output voltage. This voltage difference also falsifies the measurements of the voltage drop over the shunt resistor. For low shunt resistor values, the voltage difference $V_\text{iov}$ is the main contributor to the error current which is then given as $I_\text{err} = V_\text{iov} / R_\text{shunt}$. The error current of $\unit[2.2]{nA}$ in case of the $\unit[10]{k\Omega}$ shunt resistor corresponds to an input offset voltage of $V_\text{iov} = \unit[22]{\mu V}$, which again is in very good agreement to the data sheet's value of the input offset voltage of IC15.

\begin{figure*}[htb]
	\centering
	\includegraphics[width=1\textwidth]{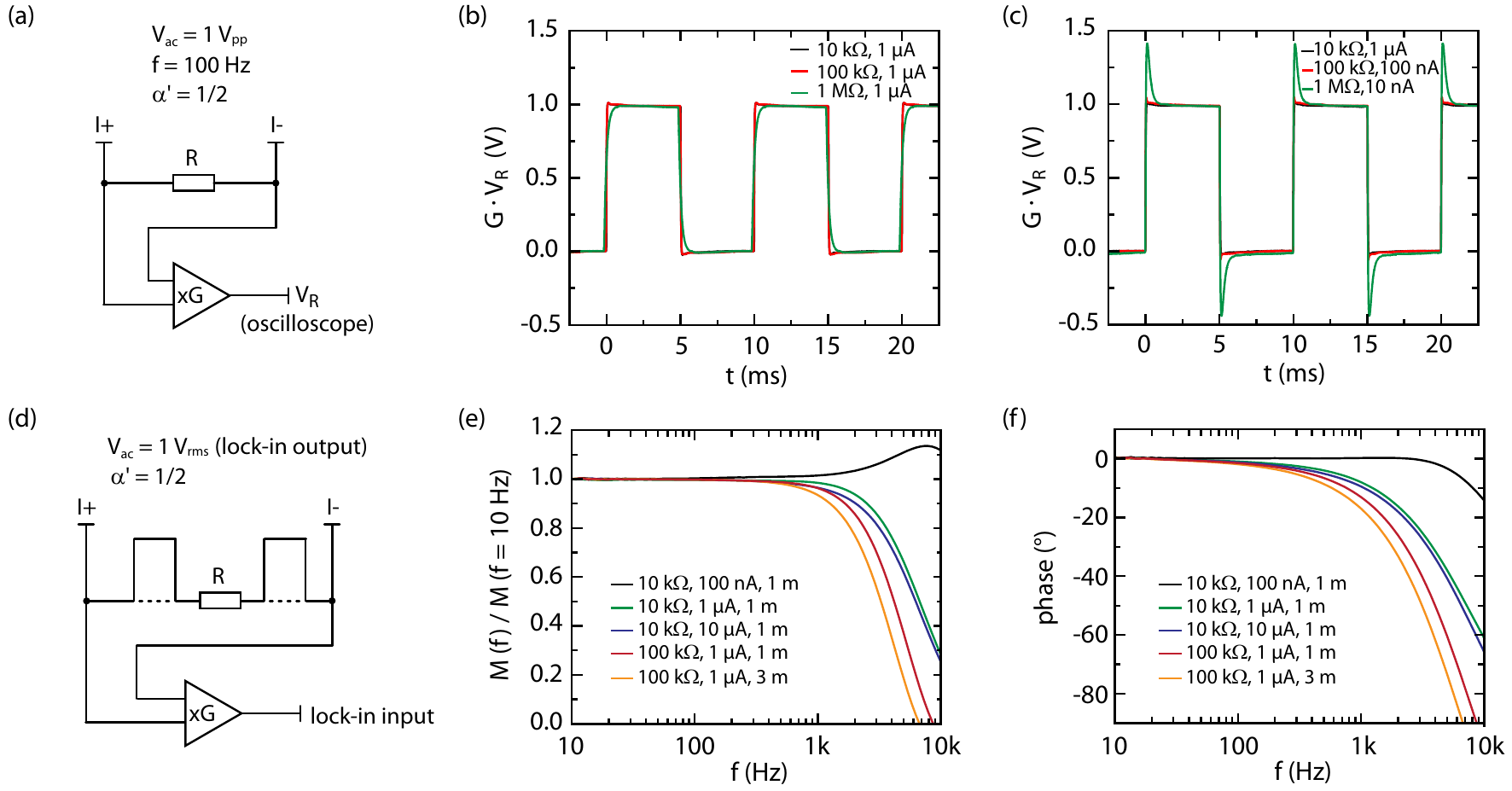}
	\caption{AC performance and bandwidth of the voltage controlled current source: a) Setup for measuring the response to rectangular pulses. The current is measured via the voltage drop over an ohmic resistor $R$ and recorded with an oscilloscope. b) and c) show corresponding measurements for different resistance values of $R$ and different current settings. d) Setup for measuring the bandwidth. The sinusoidal output voltage of a lock-in amplifier is used as a reference voltage and the voltage drop over the resistor is applied to the input of the lock-in. e) Frequency dependent change in the measured amplitude and f) the change in the phase for different values of $R$, applied currents $I_\text{DUT}$ and lengths of the used BNC cables.}
	\label{FigS5}
\end{figure*}

If identical input offset currents of $I_\text{ioc} \approx \unit[100]{pA}$ and input offset voltages of $V_\text{iov} \approx \unit[22]{\mu V}$ are to be assumed over the whole range of shunt resistor values, their contributions to the overall error current will be approximately the same in case of a shunt resistance of $\unit[100]{k\Omega}$. Now, it is important that the error currents from both sources do not necessarily have to add up, but can also compensate each other to some extent. This is the reason for the deviating shape of the curve for the $\unit[100]{k\Omega}$ shunt resistor (green curve in Fig.\,\ref{FigS4}).

For IC15 an instrumentation amplifier was chosen which has its sweet spot between input offset current and input offset voltage exactly for the one shunt resistor value which is most important for our non-local spin valve measurements: The $\unit[100]{k\Omega}$ shunt resistor that yields a current of $\unit[10]{\mu A}$ per \unit[1]{V} of reference voltage. The relative deviation in this case is very low over the whole range of $V_\text{dc}$. This is especially important if the reference voltage is eventually switched to the sinusoidal reference voltage of the lock-in amplifier, which signal periodically sweeps over the whole voltage axis of Fig.\,\ref{FigS4}. The low deviation over the whole range will therefore minimize the overall distortion of the sinusoidal signal.

If a current source with better precision in the nA range is needed, IC15 should be replaced with an instrumentation amplifier with FET-inputs. These amplifiers exhibit much lower input offset currents. Unfortunately, at the same time the input offset voltages are much higher and reach values of at least several hundreds of $\mu$V. In such a case the relative deviation is mainly given by $V_\text{iov}/V_\text{dc}$.

Another drawback of an instrumentation amplifier with FET-input is its larger temperature drift both in regard to input offset voltage and input offset currents. Instead, the amplifiers used for our current design (AD8422BRZ for the instrumentation amplifiers and OPA192 for the operational amplifiers) exhibit such low temperature drifts, that we measured a relative change of only $5\cdot10^{-4}$ in the $\unit[1]{\mu A}$ current output ($V_\text{ref} = \unit[1]{V}$, $R_\text{ref} = \unit[1]{M\Omega}$) when increasing the temperature of the current source from $\unit[25]{^\circ C}$ to $\unit[50]{^\circ C}$.

Finally, also characteristics such as noise performance, gain bandwidth product, or common-mode rejection ratio should be considered if one wishes to use another instrumentation amplifier for IC15.

\section{AC performance}
\label{ACperformance}
To investigate the AC performance of the current source, we first measure its response to rectangular pulses with a peak-to-peak amplitude of $\unit[1]{V_\text{pp}}$ and a frequency of $f=\unit[100]{Hz}$ applied as the input voltage $V_\text{ac}$ (with $\beta=1$). As a DUT we use an ohmic resistor. Therefore, the time-dependent current through the DUT can be directly measured via the voltage drop $V_\text{R}$ over the resistor $R$ (see Fig.\,\ref{FigS5}a). This voltage drop is measured with a Stanford Research SR560 low-noise amplifier (set to DC-coupling, no additional filters) which sends the amplified signal to an oscilloscope.

Fig.\,\ref{FigS5}b shows the corresponding signals for different values of $R$ for a fixed current setting of $\unit[1]{\mu A}$ applied in the on-state of the pulse ($R_\text{shunt}=\unit[1]{M \Omega}$). Instead, Fig.\,\ref{FigS5}c shows the signals for the same resistor values as in Fig.\,\ref{FigS5}b, but with decreasing currents for increasing resistance values. All signals follow the rectangular pulse shape quite reasonably, except for the case of $R=\unit[1]{M\Omega}$ and a current of \unit[10]{nA} (green curve in Fig.\,\ref{FigS5}c). The problem in this case is the RC low-pass filter which is created by the used shunt resistor for this current range ($R_\text{shunt}=\unit[100]{M\Omega}$) and the capacitance of the used BNC cables (around \unit[100]{pF} for a \unit[1]{m} long cable). The corresponding cutoff frequency of this low-pass filter is around $\unit[16]{Hz}$.

To increase the usable frequency range for currents in the lower nA range, lower shunt resistor values together with lower reference voltages $V_\text{ref}$ have to be used according to $I_\text{DUT}(t) = V_\text{ref}(t) / R_\text{shunt}$. One downside of this approach is the lower accuracy as discussed in section~\ref{DCAccuracy}. Furthermore, we also expect an increase in the noise level: As the voltage drop $R_\text{shunt} \cdot I_\text{DUT}$ over the shunt resistors will be decreased, the thermal noise of the shunt resistor (Johnson–Nyquist noise) will have an increased relative impact onto the feedback loop. This is due to the fact that the thermal noise only scales with the square root of the resistance, whereas the voltage drop that is caused by the applied current scales linearly with resistance.

Next to the RC low-pass filter, which consists out of the shunt resistor and the capacitance of the BNC cables, the overall bandwidth of the current source is also limited by both the RC low-pass filter for the reference voltage (R25 and C52 in Fig.\,\ref{FigS10}) and the components that are guaranteeing the stability of IC9 (C44, R14 and R23 in Fig.\,\ref{FigS11}). Choosing the wrong values for the latter components can easily lead to an unstable, oscillating behaviour of the whole current source. If higher frequencies are needed, these three components have to be fine-tuned to each other. For initial testing, a LTspice model of our current source design is available on Zenodo.

To investigate the bandwidth of the current source in more detail, we use the sinusoidal output voltage of a lock-in amplifier (Stanford Research SR830) with an amplitude of $V_\text{ac} = \unit[1]{V_\text{rms}}$. As a DUT an ohmic resistor $R$ is used. The current-induced voltage drop over the resistor is measured with an SR560 amplifier and is applied to the input of the lock-in (see Fig.\,\ref{FigS5}d). Both the corresponding amplitude of the signal ($M=\sqrt{X^2+Y^2}$, with $X$ being the real and $Y$ being the imaginary part) and the phase ($\varphi = \arctan (Y/X)$) are measured. Fig.\,\ref{FigS5}e depicts the frequency dependent amplitude which is normalized to the amplitude at \unit[10]{Hz}, whereas Fig.\,\ref{FigS5}f shows the corresponding phase. The measurements were made for values of $R$ and $I_\text{DUT}$ which are typical for our non-local spin valve measurements. Furthermore, the influence of the capacitance of the BNC cables is demonstrated by both using \unit[1]{m} and \unit[3]{m} long cables which are connecting the resistor to the BNC connectors $I_+$ and $I_-$ of the current source.

\begin{figure*}[htb]
	\centering
	\includegraphics[width=1\textwidth]{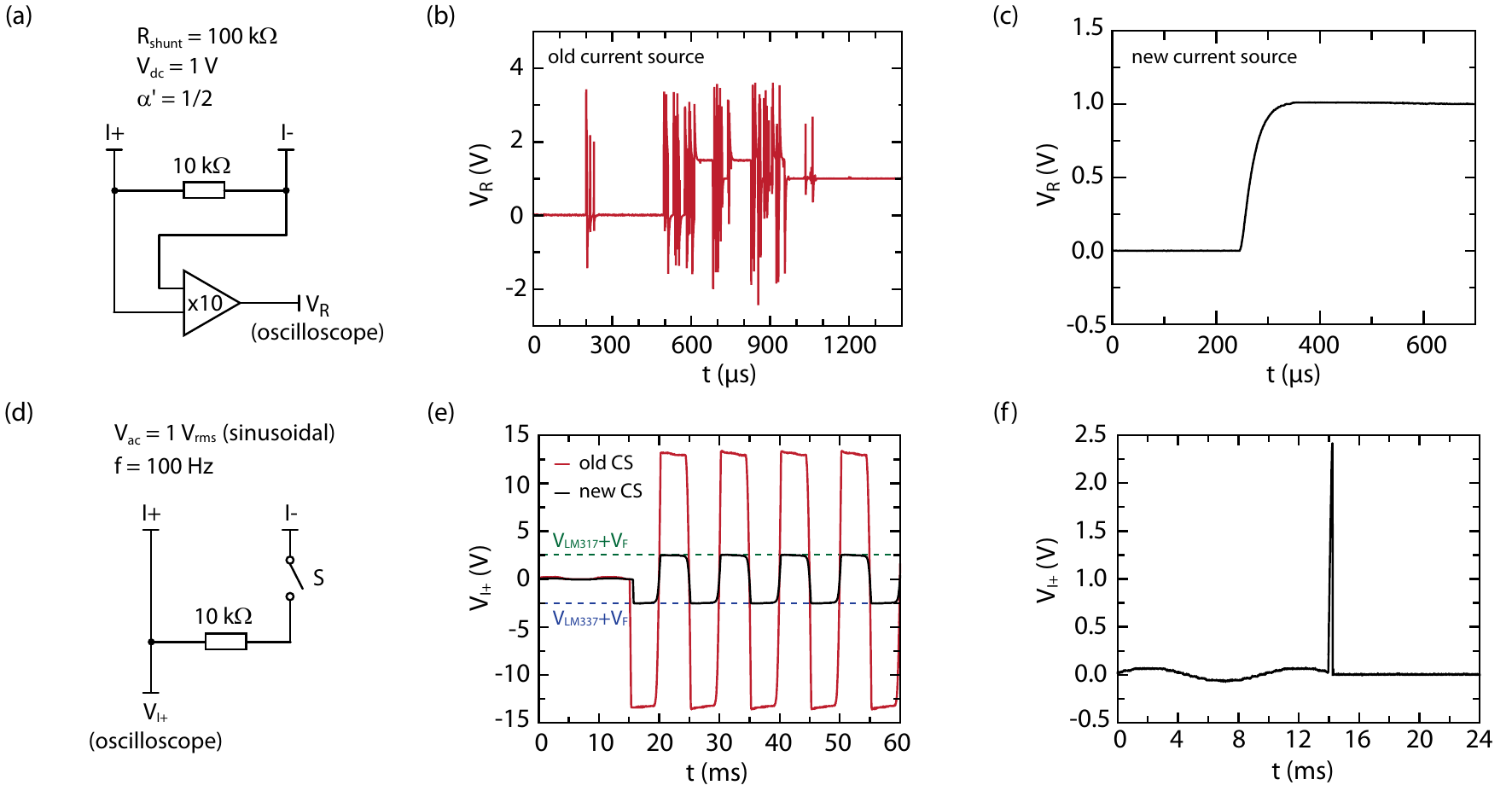}
	\caption{a) Setup for measuring the turn-on behavior of a voltage controlled current source if an external reference voltage is already applied to it. Corresponding measurements are shown in b) for the so-far used old current source and in c) for our new design, which shows a smooth transition without any overshoot or erratic behaviour. d) Setup for measuring the behavior of the current sources in the event of an interruption of the current path and e) corresponding response of both the old (red curve) and new (black curve) current source. The latter was recorded in e) in case of a deactivated and in f) in case of an activated shut-down mechanism.}
	\label{FigS6}
\end{figure*}

Figs.~\ref{FigS5}e and \ref{FigS5}f show that for frequencies $\unit[<300]{Hz}$ the change in amplitude is $<1\%$ and the measured phase is $\unit[<5]{^\circ}$. Therefore, in our lock-in based measurements we limit the applied frequencies to this frequency range to keep the absolute error small. However, we note that in many experiments not absolute values but relative changes are important and that the relative accuracy of the current source is higher. For example, in a Hanle measurement the relative change in the non-local signal with a varying magnetic field is measured. The fitted spin lifetime does not change if the Y-axis has to be rescaled by a constant factor because the applied current was different than expected. Instead, the rescaling of the Y-axis only impacts values like the spin injection efficiency.

\section{Turning the current source on/off}
\label{TurnOnOff}
It is good practice to check if the input voltages $V_\text{ac}$ and $V_\text{dc}$ (see Fig.\,\ref{FigS10}) are set to zero before turning on the current source. According to section~\ref{ReferenceVoltages} this will result in a reference voltage $V_\text{ref}(t) = 0$ applied to the non-inverting input of IC9 (see Fig.\,\ref{FigS11}). And according to section~\ref{CurrentSource} the current source will therefore try to avoid any voltage drop over the shunt resistors, hence $I_\text{DUT}(t) = V_\text{ref}(t) / R_\text{shunt} = 0$.

The on/off switch of the current source (S6 in Fig.\,\ref{FigS11}) will additionally pull the voltage at the non-inverting input of IC9 to GND if the switch is in its closed position (the resistor R25 of the low-pass filter in Fig.\,\ref{FigS10} will guarantee that the non-inverting input of IC9 is pulled to GND even if the input voltages $V_\text{ac}$ or $V_\text{dc}$ are not zero). At the same time, switch S6 also pulls the nodes "ground I+" and "ground I-" to GND. As the names suggest, the outputs of the BNC connectors labeled with "I+" and "I-" will be put to GND. Therefore the DUT, which is normally also grounded at the start of an experiment, can be safely connected to these two BNC connectors (see Fig.\,\ref{FigS11}).

As an additional safety backup, also switch S7 in Fig.\,\ref{FigS11} can be closed during the rewiring of the DUT. In its closed position, switch S7 will connect the two BNC outputs "I+" and "I-" via resistor R22, which is therefore put parallel to the DUT. If switch S6 fails to ground nodes "ground I+" and "ground I-", the connection via R22 will nevertheless close the current path between IC8 and IC14 and therefore the feedback loop can operate normally even if the DUT is removed.

Accordingly, after connecting the DUT to "I+" and "I-", first switch S6 is opened (the DUT and the reference voltage will no longer be pulled to GND). Thereafter, S7 is opened (the parallel path between connectors "I+" and "I-" will be removed). Finally, the input voltages $V_\text{ac}$ and $V_\text{dc}$ can be applied to the current source to drive the current
\begin{eqnarray}
    I_\text{DUT}(t) &=& V_\text{ref}(t) / R_\text{shunt} \\ \nonumber
    &=& \left[ \beta \cdot V_\text{ac}(t)+V_\text{dc} \right] / R_\text{shunt}
\end{eqnarray}
through the DUT (with $\beta$ being either 1 or 0.01 based on the position of S2 in Fig.\,\ref{FigS10}).

To shut the current source down, the reversed order is advised: First, decrease the amplitude of the externally applied voltages $V_\text{ac}$ and $V_\text{dc}$ to zero, then close switch S7 and, finally, close switch S6.

Of course, the question arises what will happen if this recommended procedure is not been followed. It is important to investigate this situation, as the use of current sources for experiments is sometimes discouraged because of an allegedly unstable and erratic behaviour. To illustrate this, we measured the turn-on behavior of both our old and newly designed current source via the setup shown in Fig.\,\ref{FigS6}a. Here, a $\unit[10]{k\Omega}$ resistor was used as the DUT and the voltage drop over the resistor was measured with an SR560 amplifier (10x gain, DC-coupling, no additional filters). The amplified output was then recorded with an oscilloscope (DC coupling, no averaging). For our newly designed current source the shunt resistor was set to $R_\text{shunt}=\unit[100]{k\Omega}$, switch S7 was opened and a reference voltage of $V_\text{dc}=\unit[1]{V}$ was applied. Only then, the current source was switched on by opening switch S6, resulting in an abrupt change in the reference voltage $V_\text{ref}$ applied to the non-inverting input of IC9 in Fig.\,\ref{FigS11}. In case of our old voltage controlled current source (no schematics are available), we used an analog procedure with the same current settings.

For both current sources an increase in the measured voltage from \unit[0]{V} to
\begin{eqnarray}
      V_\text{R} &=& G\cdot R \cdot I_\text{DUT}
      = G\cdot R \cdot \frac{V_\text{ref}}{R_\text{shunt}} \\ \nonumber \nonumber
      &=& 10\cdot \unit[10]{k\Omega} \cdot \frac{\unit[1]{V}}{\unit[100]{k\Omega}}
      = \unit[1]{V}
\end{eqnarray}
is expected. The actual measurements are shown in Fig.\,\ref{FigS6}b for the old current source and in Fig.\,\ref{FigS6}c for our new design. In case of the old current source, several short spikes with amplitudes significantly exceeding the final voltage of \unit[1]{V} can be observed. Instead, our new current source shows a stable turn-on behaviour. Despite the sudden increase in the reference voltage and a possible bouncing of switch S6, the desired current is reached after around $\unit[100]{\mu s}$ with a smooth transition without any overshoot or erratic behaviour.

\section{Over-voltage and current protection}
\label{OVP}
One important part in the design of a current source is its handling of either an overload condition or an outright interruption of the current path. Such an interruption can not only occur if the DUT is damaged, but also if e.g.\,the BNC cable is broken or if the wiring is incorrect due to an user error. Although the current source is unable to drive any current in these cases, it will nevertheless try to achieve it by ramping up the voltage to the highest possible value. The test setup to simulated such a situation is depicted in Fig.\,\ref{FigS6}d, whereas Fig.\,\ref{FigS6}e shows the corresponding measurements both in case of our previously used current box (red line) and our newly designed one (black line). For this measurements we use an ohmic $\unit[10]{k\Omega}$ resistor as the DUT, apply a sinusoidal reference voltage with a frequency of \unit[100]{Hz}, then interrupt the current path by opening the switch S that is connecting the DUT with the I$_-$ terminal and measure the voltage at the one side of the DUT which is connected to the I$_+$ terminal with an oscilloscope.

The switch S is opened at $t\approx\unit[15]{ms}$, which results in a sudden jump of the voltage that is applied to the $I_+$ side of the DUT. In case of the old current source (see red line in Fig.\,\ref{FigS6}e), the voltage rises almost to the supply rail of the one amplifier which drives the current (a difference to the supply rail's voltage is present due to the output voltage swing of the amplifier). For a whole half-circle of the reference signal, the voltage is pinned at this level before it abruptly jumps to the reversed polarity for the second half of the sinusoidal reference signal.

This behaviour can damage the DUT in two different ways: Although the DUT is disconnected on the $I_-$ side, the jumps by almost $\unit[25]{V}$ in a short span of time can lead to significant charging/discharging currents over the DUT due to parasitic or even intentional capacitances which are referenced to ground on the $I_-$ side of the DUT (and of course capacitances of every other connected equipment to the DUT, like e.g.\,the differential voltage amplifier which measures the non-local voltage). These charging and discharging currents in return can destroy micro- and especially nanostructures because of e.g.\,Joule heating or electromigration. The second problem arises if the DUT incorporates some kind of gated transport channel. The voltage difference between the transport channel and the gate will also undergo these sudden jumps by $\unit[25]{V}$ and may surpass the breakdown voltage of the gate dielectric.

To avoid these two possible sources of irreparable device damage, two passive protection mechanisms are incorporated into the new current source, one to prevent an over-voltage condition and one to prevent excessive charging/discharging currents over the DUT. However, there is one requirement that severely limits the implementation of such protection mechanisms: The fault currents that are due to these mechanism must be much smaller than the smallest current level which is intended to be used with the current source, i.e.\,the fault currents have to be much smaller than \unit[10]{nA} in our case. This requirement limits the protection mechanism to rudimentary ones: To prevent the over-voltage condition, the voltages at the outputs of the I$_+$ and I$_-$ terminals are clamped to the adjustable voltage rails of the LM317 and LM337 linear regulators in Fig.\,\ref{FigS9} by diodes with extremely low leakage currents (D30 to D33 in Fig.\,\ref{FigS11}; at room temperature the type 1N3595 diode only shows leakage currents of around $\unit[1]{nA}$ if the reverse bias voltage exceeds $V_\text{R} = \unit[125]{V}$).

Accordingly, the voltage level to which the signal is pinned in Fig.\,\ref{FigS6}e for the new current source is given by the adjustable voltage of the linear regulators plus the forward voltage $V_\text{F}$ of the diodes. In our design, this voltage level can be adjusted by the potentiometers VR3 and VR8 in Fig.\,\ref{FigS9} to be somewhere between \unit[2]{V} and \unit[10]{V} depending on the requirements of the measurements. If lower voltages are necessary, the voltage regulators have to be replaced by low dropout regulators (LDOs). It should always be kept in mind to include a low enough resistive load to the output of the LDOs so that the current over this load in normal operation is much larger than the current sourced or drained into these voltage rails by the clamping diodes in a fault condition.

The passive protection against too high charging/discharging currents is even more rudimentary: It consists just of high ohmic resistors between the output terminals I$_+$ or I$_-$ and the operational amplifiers IC8 and IC14 which source and drain the current through the DUT in Fig.\,\ref{FigS11}. Usually, one of the output terminals is always high ohmic because of the shunt resistor (R52 to R56 at the output of terminal I$_+$). In our old current source, the other terminal I$_-$ was instead always low ohmic. Accordingly, severe damages to previous devices (e.g.\,blown away electrodes) were almost exclusively seen at the one contact of the device which was connected to this low ohmic output. In our new design, switches S10 and S11 in Fig.\,\ref{FigS11} are in fact one single rotary switch (they are depicted separately for the sake of a clearer presentation). This rotary switch not only selects the shunt resistor value (R52 to R56), but simultaneously puts a resistor with identical value (R60 to R64) to the other output. As the resistance of the shunt resistor (R52 to R56) can be higher than the resistance of the DUT, such a symmetric placement of resistors will also push the value of the voltage divider ratio $\alpha'$, which is necessary to put the virtual ground somewhere into the DUT, towards 0.5 (see section~\ref{CurrentSource}). The manually adjustable part of the voltage divider is accordingly designed in such a way, that values around $\alpha'=0.5$ can be set more precisely than values towards 0 or 1.

The protection mechanisms described so far are designed to protect the DUT and are purely passive in nature to guarantee an immediate response time. We furthermore also incorporated an active protection mechanism in our current source, which shuts down the current source under \unit[1]{ms} after an over-voltage condition is registered. This mechanism is there to give the user an unmistakable indication that an over-voltage event has occurred and also to prevent any damage to the DUT or other connected measurement equipment by the repeated sharp voltage swings seen in Fig.\,\ref{FigS6}e.

This active protection circuit monitors the voltages at nodes "sense~I+" and "sense~I-" in Fig.\,\ref{FigS11}, i.e.\,more or less the voltages applied to the output terminals I$_+$ or I$_-$ if the small, current-induced voltage drop over the resistors R46 and R59 is neglected. These voltages are buffered by IC11 and IC16 in Fig.\,\ref{FigS13}, which have RC low-pass filters at their outputs so that the protection circuit is less likely to be triggered by very short voltage transients. Then the two monitored voltages are compared to the voltages of the LM317 and LM337 regulators ("ShutdownPos" and "ShutdownNeg" are just RC filtered signals from these regulators; see R65, R66, C68, and C69 in Fig.\,\ref{FigS13}). Therefore, it is always monitored if the voltages at nodes "sense~I+" and "sense~I-" in Fig.\,\ref{FigS11} are within the voltage range which is spanned by the two independently adjustable voltage rails. Note that the clamp diodes D30 to D33 in Fig.\,\ref{FigS11} limit the voltage at the same node to a slightly higher range because of the forward voltages of the diodes. Therefore, in case of a over-voltage condition the comparators will first start to detect this event, before the clamping diodes start to conduct at slightly higher voltages.

The comparators of IC13 and IC7 have open-collector outputs. As long as all outputs of IC13A to IC13D (plus the ones of IC7D and IC7A) are open, the $\unit[1]{M\Omega}$ resistor R8 pulls the non-inverting input of IC7B to \unit[12]{V}, which is higher than node "OverLoadPos" which is set to be roughly \unit[9]{V}. Therefore, the output of IC7B is open in this normal condition. But if one of the voltages at "sense~I+" or "sense~I-" goes either above the positive voltage rail of LM317 or below the negative voltage rail of LM337, one of the corresponding comparators IC13A to IC13D will close its output and therefore will discharge the capacitor C46 over R21. Resistor R21 is there to limit the discharge current to protect the open-collector outputs, especially if higher capacitor values are chosen for C46. The values of C46 and R8 determine the rise time of the voltage across C46, i.e.\,how long the voltage at the non-inverting input of IC7B will be below \unit[9]{V} even if the comparator, which was responsible for the discharging of C46, switches back.

The time constant given by C46 and R8 is important to guarantee the triggering of the shut-down mechanism: The open-collector output of IC7B will drive a current through the photocoupler IC2 in Fig.\,\ref{FigS14}. The triac driver output of IC2 will eventually latch, i.e.\,independently from the state of the comparators a current will flow through the two reed relays RY1 and RY2 as long as the circuit is not manually interrupted by operating the switch S1 (we use a normally-closed, momentary action switch). If the two normally-open reed relays are active, they will push nodes "ground Vref", "ground I+" and "ground I-" to ground. As the names imply, "ground I+" and "ground I-" are connected to the I$_+$ and I$_-$ terminals in Fig.\,\ref{FigS11}. Therefore, when the active protection circuit is triggered, the two terminals and with this the DUT will be put to ground. Resistors R46 and R59 in Fig.\,\ref{FigS11} will limit the current which may flow in this case. For very sensitive devices also R15 and R20 in Fig.\,\ref{FigS14} can be replaced by higher value resistors to limit the current even further.

At the same time, the reference voltage at the non-inverting input of IC9 in Fig.\,\ref{FigS11} will also be pushed near ground (see node "ground Vref"). Therefore, the current source will no longer try to drive the set current through the DUT. The reference voltage will not be exactly zero due to the voltage divider which is given by resistors R25 in Fig.\,\ref{FigS10} and R12 in Fig.\,\ref{FigS14}. The lower the value of R12, the closer the reference voltage will be near zero. But at the same time the discharge current of C52 over the reed relay RY1 will increase, potentially limiting the lifetime of the relay.

It should be noted that not only the voltages at the terminals I$_+$ and I$_-$ are monitored by the comparator circuit, but also the output voltage of IC9 (see node "V~drive" in Fig.\,\ref{FigS11} and Fig.\,\ref{FigS13}). IC7A and IC7D monitor this voltage and will trigger as soon as this voltage exceeds the range of $\unit[\pm 9]{V}$. This is an additional, redundant control mechanism: In case that the feedback loop is not stable, the output voltage of IC9 is likely to go near the supply rails of IC9 (see specifications about the output swing in the respective data sheet) and therefore trigger the shutdown circuit.

For the measurement in Fig.\,\ref{FigS6}e, the active shutdown circuit was deactivated by opening switch S1 in Fig.\,\ref{FigS14}. Instead, Fig.\,\ref{FigS6}f shows the same measurement with activated shutdown circuit. Note the zoom-in in both voltage and time axis compared to Fig.\,\ref{FigS6}e. Therefore, the sinusoidal waveform in case of the normal operation of the current source can be seen much more clearly before the switch S in Fig.\,\ref{FigS6}d is opened at $t\approx \unit[14]{ms}$. After a short voltage spike, which is limited by the clamping diodes to around \unit[2.5]{V}, the current source completely shuts down $\unit[350]{\mu s}$ after the switch S was opened. Thereafter, the voltage measured at the I$_+$ terminal is constant at \unit[0]{V}. The most limiting factor in the response time of the active shutdown circuit is the operation time of the reed relays (max.\,\unit[1]{ms} according to the data sheet).

\begin{figure*}[tb]
	\centering
	\includegraphics[width=1\linewidth]{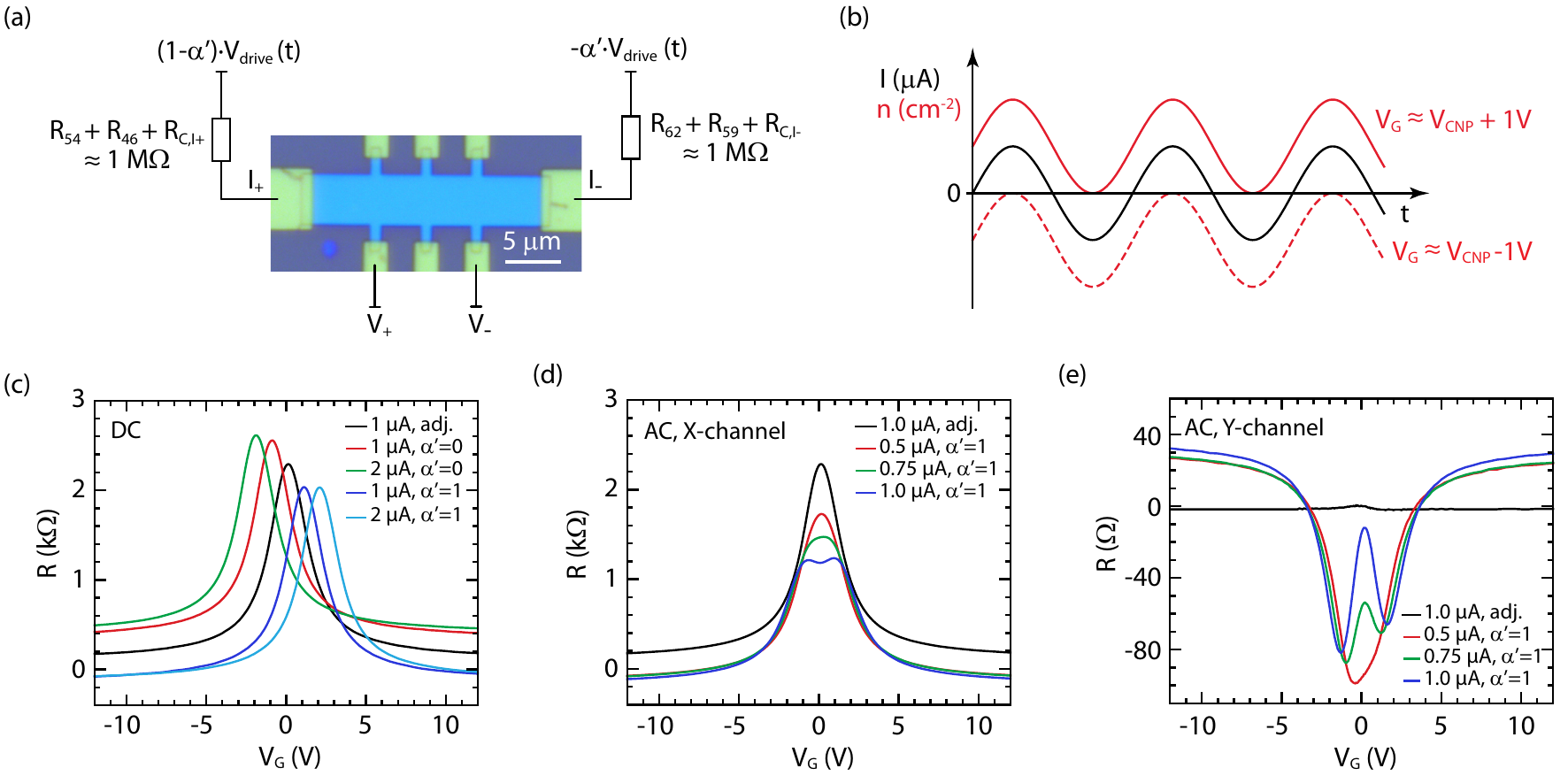}
	\caption{(a) Four-probe measurement scheme of the gate-dependent resistivity of a hBN/graphene/hBN Hall bar structure (compare to Fig.\,\ref{FigS3}c and Fig.\,\ref{FigS11}). (b) If the virtual ground is not tuned into the Hall-bar structure, any current that is driven through the whole setup (black curve) will pull the transport channel of the Hall bar to a common-mode voltage unequal to GND. This results in a change in the gate-induced charge carrier densities (red curves), because the gate voltages are referenced to GND. (c) In a DC measurement the gate-dependent position of the charge neutrality point (CNP) therefore shifts proportionally to the applied current. The direction of the shift can be inverted by changing the sign of the common-mode voltage (difference between $\alpha'=0$ and $\alpha'=1$). In addition to this electrostatic effect, also a common-mode related vertical shift can be seen in these measurements. (d) In an AC measurement the time-varying common-mode voltage results in a broadening and eventually in a splitting of the CNP towards larger currents. (e) Even in these local measurements a signal appears in the Y-channel of the lock-in as soon as common-mode related artifacts are present in the measurement.}
	\label{FigS7}
\end{figure*}

\section{Common-mode voltage induced artifacts in local measurements}
\label{VcmLocalMeasurement}
In this section we discuss how a non-zero common-mode voltage in the transport channel of a device can create artifacts even in local measurements. For this purpose we conduct measurements at room temperature on a hBN/graphene/hBN Hall bar that is fabricated on top of a Si$^{++}$/SiO$_2$ (\unit[285]{nm}) wafer (see optical image in Fig.\,\ref{FigS7}(a)). In such devices a plate capacitor model is normally used to calculate the gate-induced charge carrier density $n$:
\begin{equation}
    n = \frac{\epsilon_0 \epsilon_r}{d} \left( V_G - V_\text{cnp} \right),
\end{equation}
with the vacuum permittivity $\epsilon_0$, the relative permittivity $\epsilon_r$ of the dielectric layer, the thickness $d$ of the dielectric layer, the applied gate voltage $V_\text{G}$, and the voltage $V_\text{cnp}$ that is necessary to compensate any residual doping to tune the Fermi energy into the charge neutrality point (CNP) of graphene. However, as the gate-electric field effect depends on the voltage difference between gate and 2D material, this equation is only valid if the transport channel is on the same potential as the reference potential of the gate voltage, which is normally GND. Otherwise, the equation must include the common-mode voltage $V_\text{cm}$ of the transport channel:
\begin{equation}
\label{n(Vcm)}
    n = \frac{\epsilon_0 \epsilon_r}{d} \left( V_\text{G} - V_\text{cm} - V_\text{cnp} \right).
\end{equation}
Although our current source is designed to create a virtual ground in the transport channel, i.e.\,reducing $V_\text{cm}$ to zero, the current source can also be used to intentional increase $V_\text{cm}$ to investigate its impact on measurements.

For this purpose we use the $\unit[1]{\mu A/ V}$ range, i.e.\, a shunt resistor of $R_\text{shunt} = R_{54} = R_{62} =\unit[1]{M \Omega}$ (compare to Figs.~\ref{FigS3}c and Fig.\,\ref{FigS11} and see explanation in section~\ref{CurrentSource}). This resistance is much larger than both the contact resistances $R_\text{Ci}$ of the device and the protection resistors $R_{46}$ and $R_{59}$ at the outputs of the current source (see Fig.\,\ref{FigS11}). The total resistance between the transport channel of the Hall bar and the outputs of the operational amplifiers IC8 and IC14 therefore become  approximately $\unit[1]{M \Omega}$. The operational amplifiers IC8 and IC14 supply the voltages $(1-\alpha')\cdot V_\text{drive}(t)$ and $-\alpha'\cdot V_\text{drive}(t)$, which can be freely adjusted by the current source between $0\leq\alpha'\leq 1$ (see section~\ref{CurrentSource}).

The extreme cases of $\alpha'=0$ and $\alpha=1'$ correspond to the condition for which the virtual ground is set to the outputs of either IC8 or IC14, respectively. In either case the transport channel of the Hall bar is separated from the virtual ground by a resistance of around $\unit[1]{M \Omega}$. This means that any current $I_\text{DUT}$ that flows through the Hall bar will raise the potential of its transport channel to the common-mode voltage $\unit[1]{M \Omega} \cdot I_\text{DUT}$, which will change the electrostatic gating via equation~\ref{n(Vcm)}. This is demonstrated in the DC measurement of Fig.\,\ref{FigS7}c, where the gate-dependent position of the CNP shifts proportionally to the applied current by $\Delta V_\text{G} \approx \unit[1]{M \Omega} \cdot I_\text{DUT}$. The direction of the shift can be inverted by changing the sign of the common-mode voltage ($\alpha'=0$ vs. $\alpha'=1$).

Besides this electrostatic effect, which is of general nature, there is also a common-mode related effect that is specific to the used differential voltage amplifier, namely the vertical shift seen in the measurements of Fig.\,\ref{FigS7}c. Notably, the voltage amplifier that is used for these measurements even results in the measurement of negative resistances (see blue curves at higher gate voltages). This shift is due to a combination of all common-mode related measurement artifact that are discussed in the main manuscript. In case of DC measurements, this includes voltages that are created by leakage currents and a finite common-mode rejection ratio. Changing the differential voltage amplifier to another type with different technical specifications can change this effect significantly.

Next, we conduct AC measurements on the Hall bar device (Fig.\,\ref{FigS7}d). If the virtual ground is adjusted into the transport channel of the device, the resulting gate-dependent resistance is identical to the DC measurement with an adjusted virtual ground (black curve in Figs.\,\ref{FigS7}c and \ref{FigS7}d). Instead, a measurement with the same AC current of $\unit[1]{\mu A}$ but with a highly misaligned virtual ground ($\alpha'=1$, blue curve) not only exhibits the vertical shift with negative resistance values as discussed in the previous paragraph, but also has a highly reduced maximum resistance value and a double peak feature. This double peak is a consequence of the before-mentioned electrostatic effect: If the virtual ground is not tuned into the transport channel of the Hall bar structure, the common-mode voltage of the transport channel will change over time with the applied AC current (black curve in Fig.\,\ref{FigS7}b). For $\alpha'=1$, the common-mode voltage oscillates between $\unit[\pm 1]{\mu A} \cdot \unit[1]{M \Omega}=\unit[\pm1]{V}$. According to equation~\ref{n(Vcm)} there are now two gate voltages for which the transport channel is tuned into the CNP exactly at the times at which the current source drives a maximum current value of either $\unit[1]{\mu A}$ or $\unit[-1]{\mu A}$ through the device. The corresponding gate voltages $V_\text{G} \approx V_\text{CNP} \pm \unit[1]{V}$ (solid and dashed red curves in Fig.\,\ref{FigS7}b) explain the two peaks seen in the blue curve of Fig.\,\ref{FigS7}d.

For smaller current values (red and green curves in Fig.\,\ref{FigS7}d) or for values of $\alpha'$ closer to the case of an adjusted virtual ground condition (not shown), the time-varying common-mode voltage results in just a broadening of the CNP. We note that this purely electrostatic broadening effect towards larger currents may be easily misinterpreted in previous experiments as thermal effects induced e.g.\,by Joule-heating. Instead, adjusting $\alpha'$ in such a way that a virtual ground is created in the transport channel of the device yields $V_\text{cm}(t)=0$ independent of the applied current. To be able to freely change the current without changing the common-mode voltage makes the investigation of any current-induced effect (e.g.\,thermo-electric effects) much easier and, in particular, much less prone to misinterpretation.

\section{Procedure to adjust the virtual ground}
\label{Procedure-to-adjust-the-virtual-ground}
It is important to emphasize that even in the local measurements, that were discussed in the last section, a signal appears in the Y-channel of the lock-in as soon as common-mode related artifacts are present in the measurement (compare Figs.~\ref{FigS7}d and \ref{FigS7}e). This is similar to the non-local measurements that are discussed in the main manuscript. The occurrence of a common-mode related voltage signal in the Y-channel makes it straightforward and easy to adjust the virtual ground into the transport channel of the device during an AC measurement that is based on a lock-in: The corresponding potentiometer (VR9 or VR10) of the selected voltage divider only has to be slowly turned while monitoring the amplitude of the voltage in the Y-channel. Turning the potentiometer changes $\alpha'$ according to section~\ref{CurrentSource}. The potentiometer needs to be adjusted until the measured voltage in the Y-channel becomes minimized.

The choice of a wirewound precision potentiometer and the special design considerations of the voltage divider (see Fig.\,\ref{FigS3}) make it possible to safely operate VR9 and VR10 while the current source is in operation. We verified with oscilloscope measurements that the turning of the potentiometers is changing the measured signals in a continuous way without any observable glitches or voltage spikes. Instead, switches S3, S4, S5, S8, or S9 should only be operated if the current source is shut down.

Of course, another adjustment procedure for the virtual ground is to use one contact of the device to probe the potential of the transport channel against ground. Minimizing this potential difference by turning the potentiometer allows an adjustment also in case of DC measurements. However, we note that for DC measurements the input offset voltages and input bias currents of the voltmeter also has to be taken into account. Even if the reading of the voltmeter is zero, the potential of the transport channel can differ from ground in the worst case by the input offset voltage of the voltmeter plus its input bias current times the contact resistance to the transport channel.

\section{Applications and limitations of the current source}
\label{Applications-Limitations}
As explained in the main manuscript, the current source is extremely effective in reducing all spurious charge signals that are directly caused by a common-mode voltage inside the non-local part of the device. This includes voltage signals that are caused by 1.) leakage currents to ground ($V_\text{L}$), 2.) a finite common-mode rejection ratio ($V_\text{CMRR}$), or 3.) charging currents of capacitances ($V_\text{CC}$). Whereas the latter signal is only relevant in AC measurements (in DC measurements the impedance of capacitors is only given by their insulation resistance, which in turn contributes to $V_\text{L}$), the other two spurious signals also occur in DC measurements.

The possibility to create a virtual ground in the transport channel of a device (both for DC and low-frequency AC currents) can also have great benefits in experiments in which the electrostatic surface potential plays a role. This includes e.g.\,Kelvin probe force microscopy or electron microscopy experiments.

However, our current source only has a minor influence on the measurement artifacts that are caused by either current spreading ($V_\text{CS}$), thermo-electric voltages ($V_\text{T}$), input bias currents ($V_\text{IBC}$), or crosstalk and interference signals ($V_\text{CI}$) as mentioned in the main manuscript. This is due to the fact that these four signals are quite insensitive to changes of the common-mode voltage in the transport channel.

The current spreading effect, e.g., depends on the conductivity tensor within the transport channel and on the spatial injection and extraction profiles of charge carriers into and out of the transport channel (see references given in the main manuscript). As the leakage currents that are induced by the common-mode voltage are normally much smaller than the current that is driven between source and drain contacts in our graphene-based spin-valve devices, the overall spatial injection and extraction profiles of charge carriers only change marginally with the common-mode voltage. And the only way the conductivity tensor can change with the common-mode voltage is the effect that was discussed in section~\ref{VcmLocalMeasurement}, namely that changing the common-mode voltage in the transport channel also changes the potential difference to a gate and, therefore, the gate-induced charge carrier densities. However, we found that both effects on the current spreading are almost negligible in our graphene-based spin valve devices. For example, adjusting the virtual ground into the transport channel of our devices only effects the zeroth order of the background voltage in Hanle measurement, whereas the terms that depend linearly or quadratically on the magnetic field are almost unaffected. As demonstrated in the measurements shown in Fig.\,5 of the main manuscript, the reduction in the zeroth order can be explained by a suppression of $V_\text{L}$, $V_\text{CMRR}$, and $V_\text{CC}$.

Although common-mode-voltage-induced leakage currents may contribute to thermal voltages, we have no evidence that this effect is relevant in our graphene-based spin valve devices. This is most likely due to the fact that these leakage currents are much smaller than the charge current that we apply in the injection circuit. However, we cannot exclude the possibility that in other materials and device structures, these leakage currents may indeed contribute to measurable thermal voltages.

It is also not a-priori clear, if the input bias currents in a given measurement setup have a relevant dependence on the common-mode voltage. Unfortunately, manufacturers of ready-to-use differential voltage amplifiers often provide only sparse information on the input bias currents of their equipment (measurement procedures for the input bias currents are given in section~\ref{PerformanceTests}). Instead, we can take a look at instrumentation amplifiers that are often used for low-cost, custom-made differential amplifiers. Examples include the instrumentation amplifiers AD8429, AD8422, and AD8220 from Analog Devices or INA826, INA818, and INA121 from Texas Instruments. We have chosen these amplifiers as all data sheets include graphs of the input bias current as a function of the common-mode voltage. Additionally, these amplifiers demonstrate differences of both the overall amplitude and the common-mode voltage dependence of the input bias current between various types of amplifiers (lower bias currents are normally achieved at the expense of higher input offset voltages and/or noise levels, therefore there are different kinds of amplifiers that are custom-tailored to different applications, see also comments in section~\ref{DCAccuracy}). However, in our experience the change of the common-mode voltage has a rather negligible impact on the overall bias current of state-of-the-art differential amplifiers. In most cases, the input bias current can be rather considered as a DC current that is not affected by the applied voltage to the inputs of the amplifier. Therefore, measurement artifacts that are caused by input bias currents are normally only seen in DC measurements, whereas they are not seen in lock-in based measurement techniques.

\section{Final notes}
\label{FinalNotes}
\subsection{Minor technical notes}
Because of the long rise time of the voltage at the non-inverting input of IC7B in Fig.\,\ref{FigS13}, which is caused by the time constant of R8 and C46, the active shutdown circuit should always trigger when the current source is powered up. If the LED D20 in Fig.\,\ref{FigS14} should not light right after switching on the main power, the whole current source should be checked.

For the switches S10 and S11 in Fig.\,\ref{FigS11} a double pole, 5 throw, shorting style (make before break) rotary switch is used. As long as a DUT is connected, the shorting style guarantees that there is always a conducting path between IC8 and IC14, even if the rotary switch is operated during an actual measurement. We strongly advise against this! Always switch off the current source before changing the shunt resistor, i.e.\,the current range. The shorting style switch is just a safety feature that minimizes the risk that the current source goes erratic because of a missing feedback loop for IC9 during an accidental operation of the rotary switch.

For S2 to S11 gold plated switches are used to guarantee low ohmic connections even for small currents and logic level voltages.

\subsection{Use of dummy devices}
We advise the fabrication of dummy devices that can be used to investigate the non-local charge signals in a specific setup and under specific measurement conditions. In case of non-local spin valve devices, all charge signals that are discussed in the main manuscript can be investigated by just replacing the ferromagnetic metal of the electrodes with a non-ferromagnetic alternative.

We are also using much more rugged and long-term stable dummy devices that are built from SMD resistors. These devices follow the equivalent circuit shown in Fig.\,2b of the main manuscript with the seven resistors $R_\text{Ci}$ and $R_\text{Mi}$. The advantage of such a device is the definite absence of any spin-related signal, this includes even small signals like the spin Hall effect that may be present in spin-valve devices with non-ferromagnetic electrodes. The disadvantage is that certain spurious non-local signals like the ones caused by current spreading or thermoelectric voltages are either strongly diminished or even completely suppressed because of the macroscopic dimensions of such a device.

\subsection{Oscilloscope measurements}
We want to emphasize the importance to check all signals within a measurement setup with an oscilloscope on a regular basis. This not only includes the actual, non-local signal but also e.g.\,the current in the injection circuit by measuring the voltage over a shunt resistor. This is due to the fact that many problems can be masked by the data processing in a digital voltmeter or lock-in amplifier. Instead, time-resolved measurements with an oscilloscope can e.g.\,reveal instabilities or oscillations of the current source, coupled interference signals, or distortions of the intended sinusoidal signal due to bandwidth or slew rate limitations. All these phenomena can cause unexpected artifacts in a measurement.

\subsection{Performance tests of a measurement setup}
\label{PerformanceTests}
Depending on the type and the condition of the used equipment, measurement setups can differ considerably from one another in specifications such as common-mode rejection ratios, offset voltages, or input bias currents. Since these quantities may deteriorate over time due to damage to the equipment, we recommend to measure these parameters regularly. Guidelines for such tests can be found in manuals, application notes, or tutorials from companies either manufacturing measurement equipment or working in the semiconductor industry (see e.g.\,the manuals of the SR830 lock-in or the SR560 voltage amplifier from Stanford Research Systems, the tutorials MT-038 and MT-042 from Analog Devices, or the "Low Level Measurements Handbook" from Keithley; all of these documents are freely available for download).

\begin{figure}[tb]
	\centering
	\includegraphics[width=0.5\textwidth]{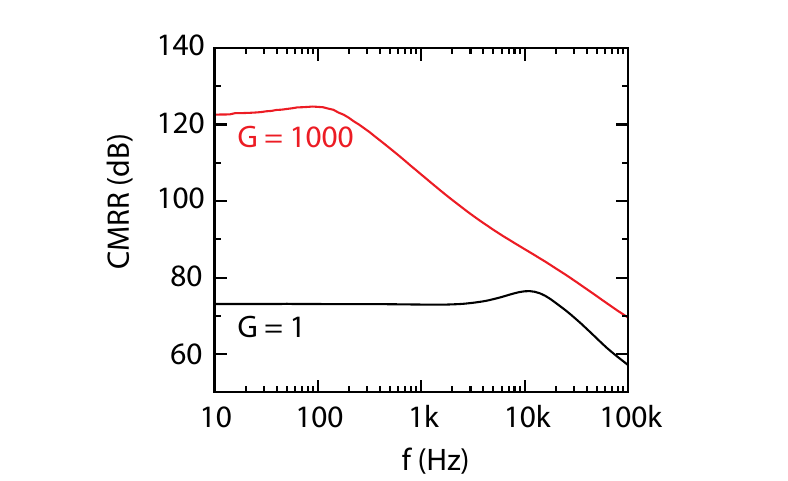}
	\caption{Measurement of the frequency dependent common-mode rejection ratio (CMRR) of the combined system of a SR560 differential voltage amplifier and a SR830 lock-in amplifier for two different gain settings of the SR560.}
	\label{FigS8}
\end{figure}

Based on the general information in the aforementioned documents, we adapted the following specific test procedures in case of our setup (i.e.\,a SR560 differential voltage amplifier that is connected to the input of a SR830 lock-in amplifier, compare to Fig.\,1a of the main manuscript). To measure the common-mode rejection ratio, we connect the output of the lock-in directly to both inputs of the differential amplifier (called input $V_\text{input}^{+}$ and $V_\text{input}^{-}$ in the following) via equally long cables and a BNC T-connector. Applying a sinusoidal output signal with an amplitude of $V_\text{ext}=\unit[100]{mV}$ and varying frequency, we measure the differential voltage output of the SR560 with the lock-in. As explained in the main manuscript, the output voltage is given as:
\begin{equation}
    V_\text{out}^\text{real} = G_\text{diff}\left( V_\text{input}^{+} - V_\text{input}^{-} \right) + G_\text{cm} \frac{\left( V_\text{input}^{+} + V_\text{input}^{-} \right)}{2}.
\end{equation}
As $V_\text{input}^{+}=V_\text{input}^{-}=V_\text{ext}$, the CMRR can be directly calculated from the measured lock-in signal $V_\text{out}^\text{real}$ via:
\begin{equation}
CMRR\left[\text{dB}\right] = 20\cdot\log\left(\frac{V_\text{ext}}{V_\text{out}^\text{real}}\cdot G_\text{diff}\right).
\end{equation}
A corresponding frequency-dependent measurement of the CMRR for two gain settings ($G_\text{diff}=1$ and $G_\text{diff}=1000$) is shown in Fig.\,\ref{FigS8}.

To measure the input bias currents of the SR560, we first set the measurement mode from differential to single-ended (i.e.\, the SR560 does not measure the voltage difference between both inputs, but the voltage difference of one of the inputs to ground). In the next step we put a shorting BNC connector cap onto the corresponding input. By doing this the input bias current has an extremely low-ohmic connection to ground and therefore does not affect the following measurement. However, even in this configuration the output voltage of the SR560 won't be perfectly zero because of input offset voltages. This input offset voltage is then measured with an Agilent 34410A multimeter. In the final step we remove the shorting BNC cap. The input bias current now has to flow over the $\unit[100]{M\Omega}$ resistor that is connecting the input to ground inside the SR560 (this is the DC part of the equipment's input impedance). This results in an ohmic voltage drop on top of the input offset voltage. Therefore, the voltage difference between the two measurements with and without the shorting BNC cap divided by the input resistance yields the input bias current of the corresponding input (note that for very low input bias currents this voltage difference might only be visible at higher gain settings).

\subsection{Replication of the current source}
In case the current source is to be replicated, we now give some ideas on what could be changed compared to our version.
\begin{itemize}
\item A change in the power supply (especially lowering the minimum voltage levels of the rails that are used for clamping) as discussed in sections~\ref{PowerSupply} and \ref{OVP}.
\item Choosing other instrumentation amplifiers to improve the accuracy of the current source for either higher or lower current ranges as discussed in section~\ref{DCAccuracy} (right now it is optimized for the $\unit[10]{\mu A}$ range).
\item Possible improvements in the bandwidth by fine-tuning the feedback loops as discussed in section~\ref{ACperformance}.
\item Adding feedback capacitors parallel to resistors like R4 or R16 in Fig.\,\ref{FigS10} to reduce the effective bandwidth of the corresponding amplifiers, making the whole circuit even more resistant against high frequency interference signals or instabilities.
\item Using reed relays to route the signals directly on the PCB instead of passing the signals back and forth to switches on the front panel (see Fig.\,\ref{FigS1}).
\item Replacing the manual potentiometers (VR9 and VR10 in Fig.\,\ref{FigS12}) with dual-supply digital potentiometers to allow a remote-controlled or even automatic adjustment of the virtual ground.
\end{itemize}

\clearpage

\pdfpageattr\expandafter{\the\pdfpageattr/Rotate 90}

\begin{turnpage}

\begin{figure*}[p]
	\centering
	\includegraphics[width=1\linewidth]{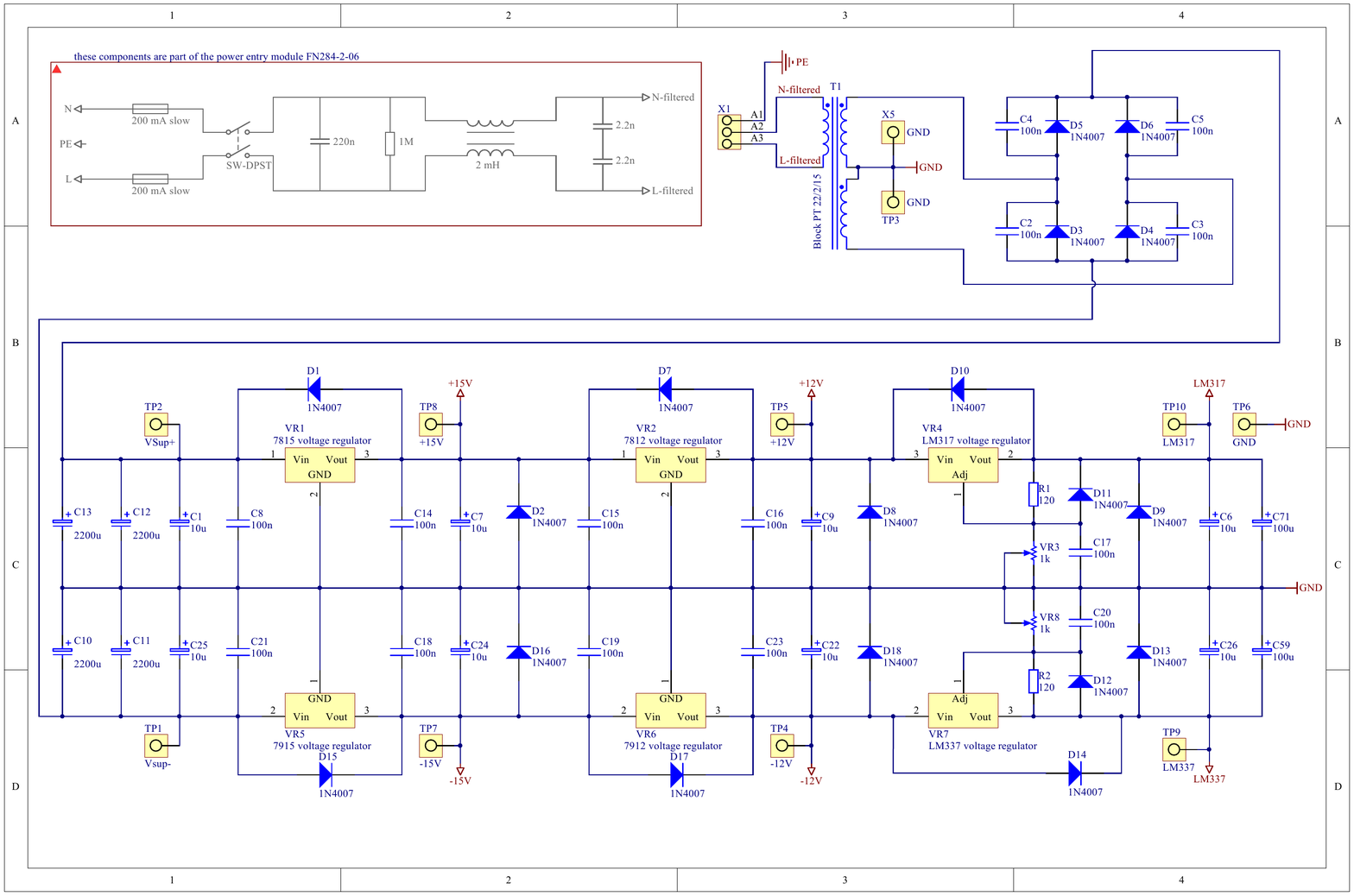}
	\caption{Schematic: Power supply.}
	\label{FigS9}
\end{figure*}

\begin{figure*}[p]
	\centering
	\includegraphics[width=1\linewidth]{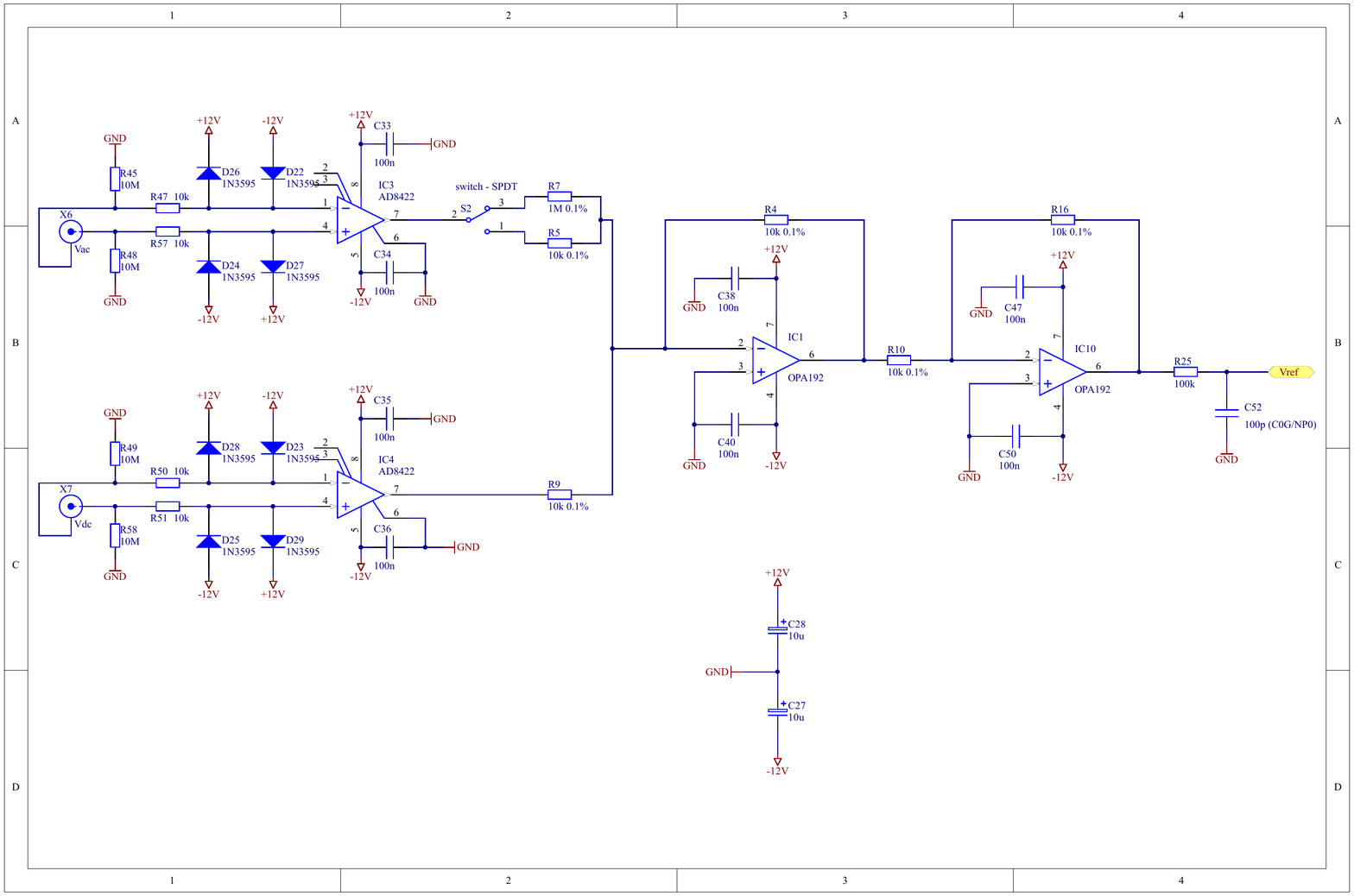}
	\caption{Schematic: Reference voltages.}
	\label{FigS10}
\end{figure*}

\begin{figure*}[p]
	\centering
	\includegraphics[width=1\linewidth]{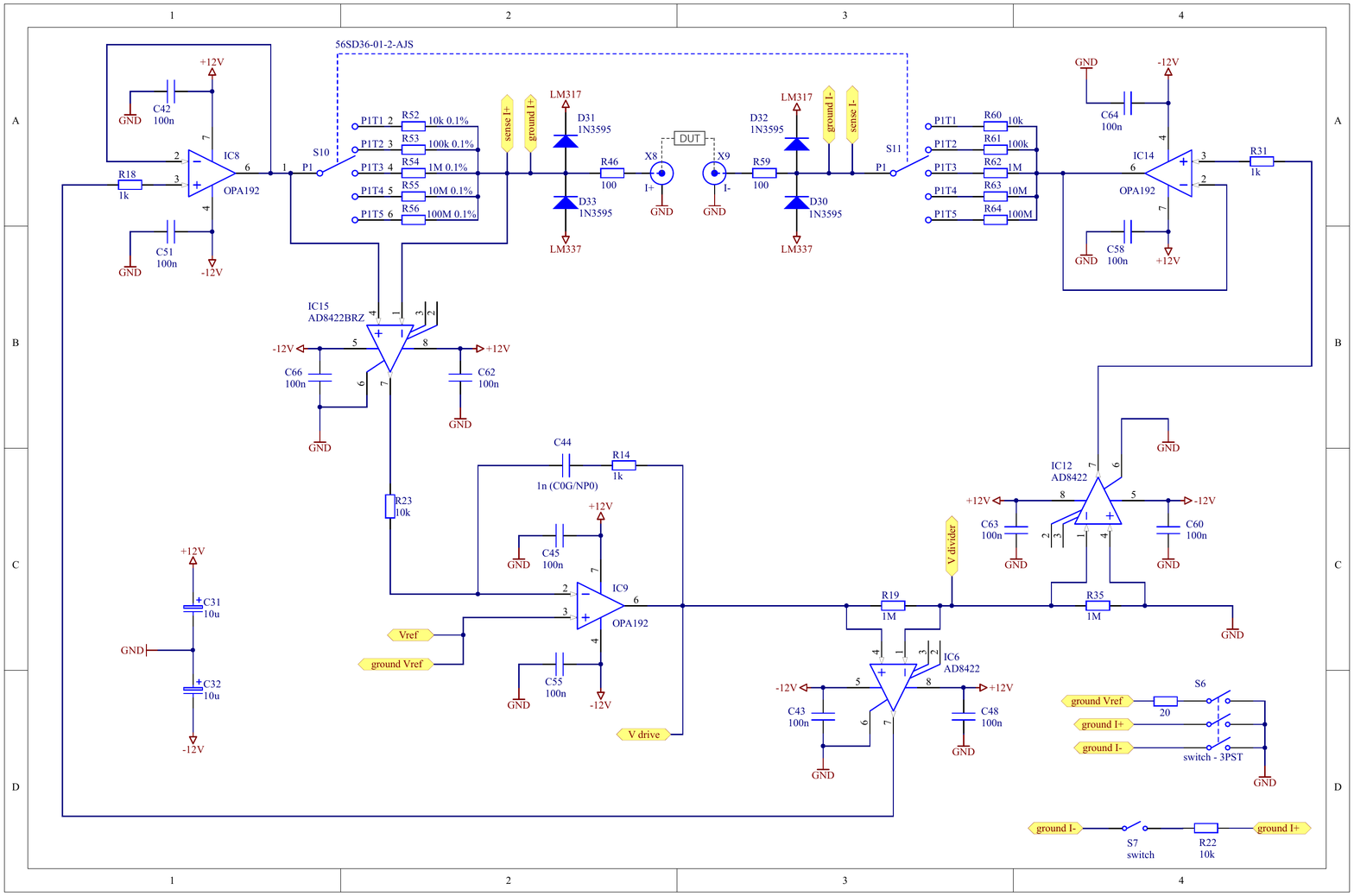}
	\caption{Schematic: Current source.}
	\label{FigS11}
\end{figure*}

\begin{figure*}[p]
	\centering
	\includegraphics[width=1\linewidth]{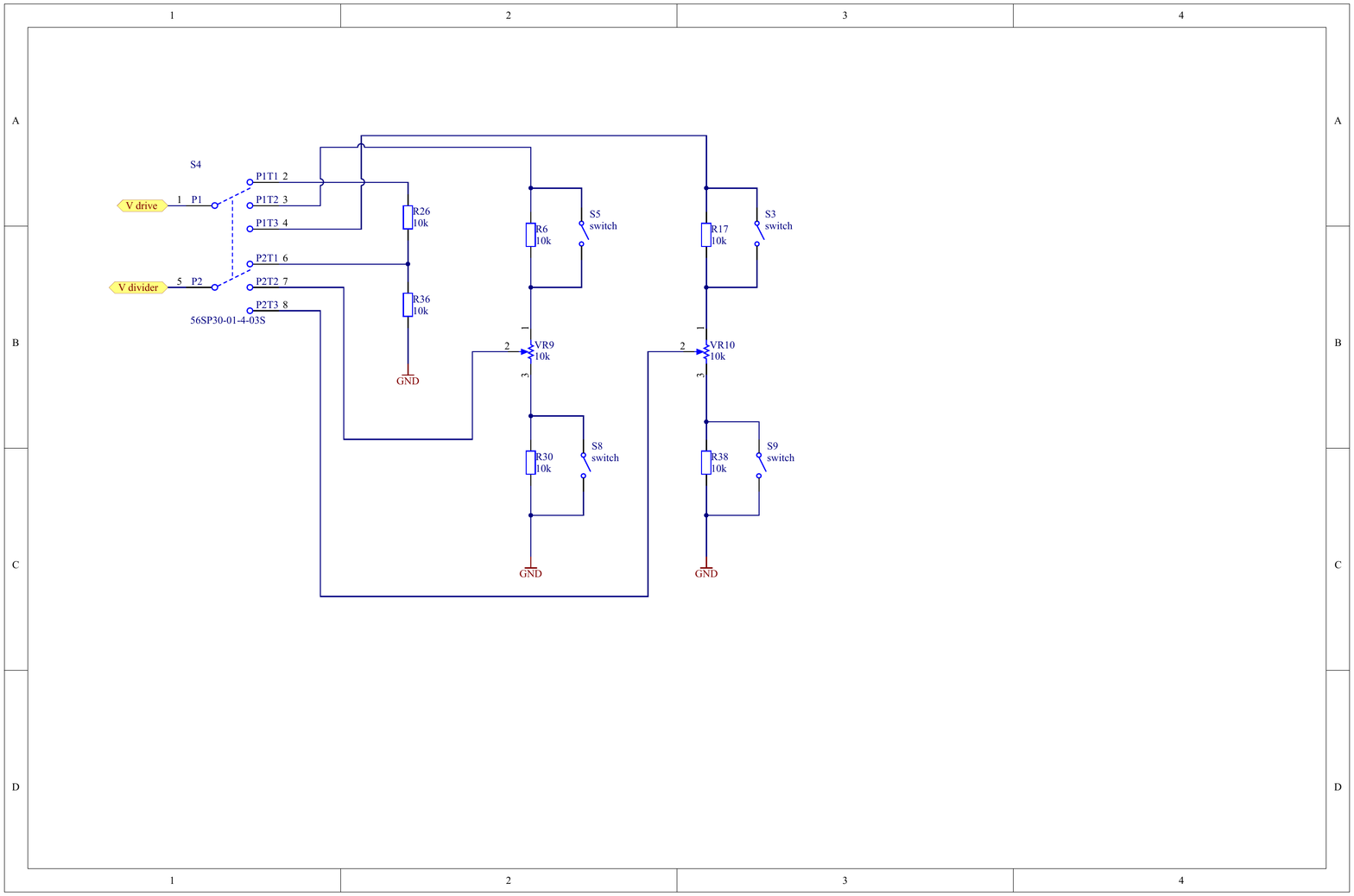}
	\caption{Schematic: Voltage divider.}
	\label{FigS12}
\end{figure*}

\begin{figure*}[p]
	\centering
	\includegraphics[width=1\linewidth]{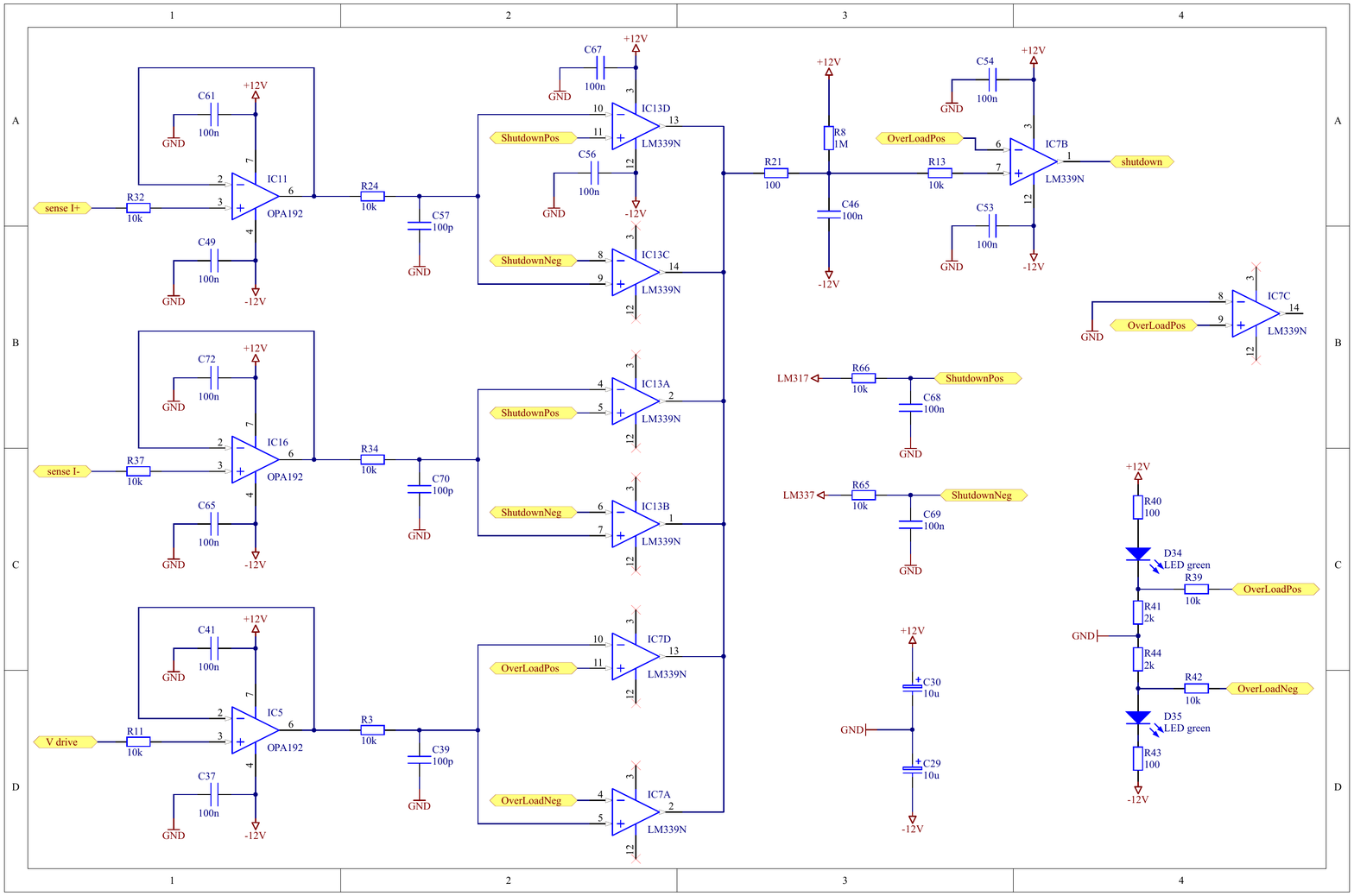}
	\caption{Schematic: Comparators.}
	\label{FigS13}
\end{figure*}

\begin{figure*}[p]
	\centering
	\includegraphics[width=1\linewidth]{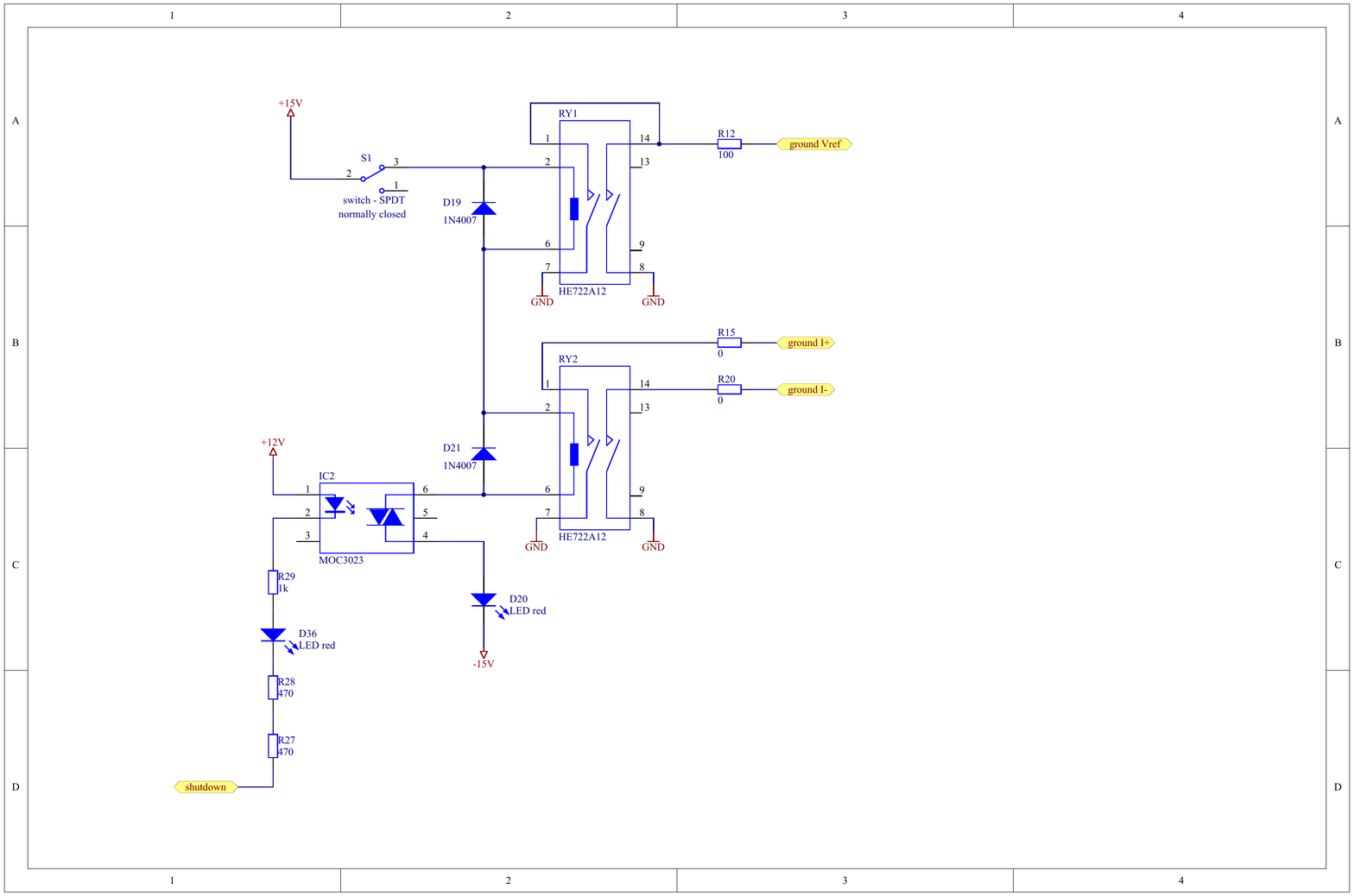}
	\caption{Schematic: Shutdown circuit.}
	\label{FigS14}
\end{figure*}

\end{turnpage}